\documentclass[twocolumn]{aastex63}
\usepackage{threeparttable}
\usepackage{amssymb}
\usepackage{graphicx}
\usepackage{CJKutf8}
\usepackage{natbib}
\usepackage{textcomp}
\usepackage{gensymb}
\usepackage{amsmath}
\usepackage{url}
\usepackage{hyperref}

\newcommand{\dn}{$\mbox{D}_n\mbox{4000}$}
\newcommand{\ewhd}{$\mbox{EW(H}\delta\mbox{)}$}
\newcommand{\ewha}{$\mbox{EW(H}\alpha\mbox{)}_{em}$}
\newcommand{\niiha}{[\ion{N}{2}]/H$\alpha$}
\newcommand{\oiiihb}{[\ion{O}{3}]/H$\beta$}

\turnoffeditone

\begin{document}
%\title{Post-starburst galaxies: a view of local analogs with the MaNGA survey}
\title{Searching for local counterparts of high-redshift post-starburst galaxies in integral field unit spectroscopic surveys of nearby galaxies}
%\correspondingauthor{Po-Feng Wu}
%\email{pofeng@mpia.de}

\author{Po-Feng Wu \begin{CJK*}{UTF8}{bkai}(吳柏鋒)\end{CJK*}}
\altaffiliation{EACOA Fellow}
\affiliation{National Astronomical Observatory of Japan, Osawa 2-21-1, Mitaka, Tokyo 181-8588, Japan}

\correspondingauthor{Po-Feng Wu}
\email{pofeng.wu@nao.ac.jp}

\begin{abstract}	
	Searching in the MaNGA IFU survey, I identify 9 galaxies that have strong Balmer absorption lines and weak nebular emission lines measured from the spectra integrated over the entire IFUs. The spectral features measured from the bulk of the stellar light make these galaxies local analogs of high-redshift spectroscopically-selected post-starburst galaxies, thus, proxies to understand the mechanisms producing post-starburst galaxies at high-redshifts. I present the distributions of absorption-line indices and emission-line strengths, as well as stellar kinematics of these local post-starburst galaxies. Almost all local post-starburst galaxies have central compact emission-line regions at the central $<1$~kpc, mostly powered by weak star-formation activities. The age-sensitive absorption line indices \ewhd\ and \dn\ indicate that the stellar populations at the outskirts are older. Toy stellar population synthesis models suggest that the entire galaxies are experiencing a rapid decline of star formation with residual star-formation activities at the centers. These features demand that most post-starburst galaxies are the aftermath of highly dissipative processes that drive gas into centers, invoke centrally-concentrated star formation, then quench the galaxies. Meanwhiles, when measurable, post-starburst galaxies have the directions of maximum stellar velocity gradients align with photometric major axes, which suggest against major mergers being the principal driving mechanism, while gas-rich minor mergers are plausible. While directly obtaining the same quality of spatially-resolved spectra of high-redshift post-starburst galaxies is very difficult, finding proper local counterparts provides an alternative to understand quenching processes in the distant Universe. 
\end{abstract}

%\keywords{galaxies:}

\section{Introduction}

Large extragalactic surveys have revealed that there are two general categories of galaxies; blue, morphologically disk galaxies that are forming new stars in-situ and red, predominantly spheroid galaxies that are quiescent in star formation activities \citep{str01, bal04, bel04}. The number density and the total stellar mass density of quiescent galaxies increase with time since $z\sim4$ \citep{bel04,fab07,ilb13,muz13b}, indicating that at any cosmic time, a fraction of star-forming galaxies stop forming stars and join the quiescent population. 

Understanding the suppression of star-forming activities in galaxies has been one of the hotly debated topic in galaxy evolution for decades. Analysis of the star formation histories (SFHs) and structures of galaxies reveal that there are multiple mechanisms at work moving galaxies from the star-forming to the quiescent population at a wide range of timescles \citep{mart07,sch14}. One type of process suppresses star formation slowly without necessarily changing the structures of galaxies. On the other hand, some galaxies had transformed both the structures and star formation activities within a short period of time \citep{sch14,wu18a,nog18}.  

The rapid truncation of star formation, or quenching, leaves unique imprints in stellar populations of galaxies. A few hundred Myr after the quenching event, the galaxy light is dominated by A-type stars. The galaxy has strong Balmer absorption lines, weak emission lines from star-forming regions, and an overall spectral energy distribution (SED) similar to A-type stars. Traditionally, these galaxies are identified spectroscopically and referred as post-starburst galaxies \citep{dre83,bal99,dre99}. 

%** Maybe focus more on local, saying that if we select from gloabal properties, we don't resolve it. If we can resolve it. we selected by local properties and not sure about whether the global properties. Then, link to high-z studies. Say that more post-starburst galaxies are at high-z, but unresolved. 

Post-starburst galaxies are rare objects in the local universe, consist of only $\sim0.1\%$ of the local galaxy populations \citep{bla04,got05,got07}. However, the fraction increases sharply along with redshifts: $\sim10$ times higher at $z\sim1$ \citep{yan09,ver10,row18a} and they may contribute to a nonnegligilble fraction of newly formed quiescent galaxies in the early universe \citep{wil16,bel19}. Understanding the properties of post-starburst galaxies and the mechanisms that produce them are thus crucial building blocks in high-redshift galaxy evolution studies. 

Observation on high-redshift post-starburst galaxies have put some constraints on their formation mechanisms. Many studies on field post-starburst galaxies at $z\gtrsim0.5$ found them having smaller half-light radii ($R_e$) in restframe optical wavlengths than quiescent galaxies of similar stellar masses \citep{whi12,bel15,yan16,alm17,wu18a,bel19}, contrary to the typical correlation between the half-light radii and the stellar ages of quiescent galaxies \citep{wu18a}. They also have flatter color gradient than older quiescent galaxies \citep{sue20}. The light and color profiles can be explained by a centrally-concentrated star-formation activity happend before quenching \citep{wu20}. Simulations show that highly dissipative processes induced by galaxy mergers and interactions are able to drive gas into galaxies centers and invoke compact starbursts then quenching \citep{bek05,wil09,sny11,zhe20}. At the same time, the merger simulations also make predictions on the properties of the remenents including the stellar age gradients and stellar kinematics \citep{bek05,zhe20}. In principle, these predictions can be tested and verified by the spatially-resolved continuum spectra of post-starburst galaxies. Nevertheless, collecting spatially-resolved continuum spectra for galaxies beyond the nearby Universe is extremely difficult: long integration time on large telescopes with good seeing \citep{deu20,set20} and preferentially with the assist of adaptive optics or gravitational lenses to resolve the galaxies. 
Furthermore, the principal driving mechanisms likely depend on both the stellar mass and the environment \citep{wu14,mal18,mat20,set20}. Collecting such data on a sample covering wide range of physical properties would be pratically infeasible. 

%\citep[see][for different results at lower-mass and in galaxy clusters]{mal18,mat20}
%which is likely due to younger stellar populations at the centers \citep{deu20}. The light and color profiles 

On the other hand, spatially-resolved continuum spectra of post-starburst galaxies in the nearby Unvierse are within the reach of current observing facilities. Early works used long-slit and integral field unit (IFU) spectrographs to follow up post-starburst galaxies identified by fiber spectra \citep{pra05,got08a,got08b,pra10,pra13}. In recent years, IFU surveys of nearby galaxies simplify the process. The sample sizes are large enough to search for rare objects like post-starburst galaxies and spatially-resolved spectra are readily available. \citet{row18b,che19} searched the Mapping Nearby Galaxies at Apache Point Observatory (MaNGA) survey \citep{bun15} for local post-starburst features and found that off-center post-starburst features are common. 
However, these local `post-starburst galaxies', either identified through fiber spectrographs or galaxies with localized post-starburst features, are likely not comparable to high-redshift post-starburst galaxies in the literature because of very different aperture sizes used to identify the post-starburst features. These studies reveal the processes that truncate star formation locally, but not necessarily globally. 

Instead, in this paper, I search post-starburst galaxies in the MaNGA survey based on the spectra integrated over the entire IFU plates. The large covering fractions are more similar to high-redshift observations. The sample should inform us about the global truncation of star formation in galaxies. Or at least, they will reveal empirically the possible routes to produce post-starburst galaxies identified by high-redshift spectroscopic observations. 

I describe the data and sample selection in Sec.~\ref{sec:data}. The analysis of IFU spectra is in Sec.~\ref{sec:result}. I then discuss the implication on the quenching mechanisms and high-redshift observations in Sec.~\ref{sec:dis} and summarize the work in Sec.~\ref{sec:sum}.

%Spatially-resolved continuum spectroscopy is powerful to study the stellar kinematics to test the merger hypothesis and unambiguously identify the central, likely compact, young stellar populations. However, such observations is extremely challenging with current facilities due to the small angular sizes and faintness of high-$z$ post-starburst galaxies. Potential age gradients, if exist, are heavily smeared out by the seeing and may also be buried in noises \citep{pra05,pra10,deu20}. 

%On the other hand, spatially-resolved continuum spectroscopy is available for post-starburst galaxies at lower redshifts. 

%Galaxy mergers and interactions are strong candidate mechanisms. In theory, the disturbance can induce gravitational instability, make gas lose angular momentum then collapse rapidly to the centers of the potential wells, induce intensive star formation activities that consum the gas quickly, and trigger feedback from stellar winds and active galatic nuclea (AGN) that further suppress star formation \citep{spr05,hop06,wil09,sny11}. One of the direct predictions from this scenario is that the centers post-starburst galaxies should be populated by young stars. 

\section{Data}
\label{sec:data}

\subsection{The MaNGA survey}

The MaNGA survey is an IFU survey of nearby galaxies as part of the 4th generation of of the Sloan Digital Sky Survey \citep[SDSS-IV,][]{bla17}. MaNGA takes spectral observations of each galaxies using fiber-bundle IFUs in the 3,600--10,000\AA\ range at a spectral resolution of $R\sim2000$. The bundles cover hexagonal regions of diameters ranging from 12" to 32" \citep{bun15}. I use the data release DR15 \citep{agu19}, which contains 4076 individual galaxies at $z < 0.15$ with a median $z\sim0.04$. 

The data is processed by the MaNGA Data Reduction Pipeline \citep[DRP,][]{law16}, which produces sky-subtracted spectrophotometrically calibrated spectra and rectified three-dimentional data cubes with a spatial pixel scale of 0.5" pixel$^{-1}$. The median reconstructed angular resolution is $\sim2.5"$ FWHM and the median spectral resolution is $\sim72$ km s$^{-1}$. The reduced data is then feed to the MaNGA Data Analysis Pipeline \citep[DAP,][]{belf19,wes19} to produce various measurements from the spectra, such as kinematics, emission line properties, and spectral line indices. The DAP use the Penalized Pixel-Fitting method \citep[pPXF,][]{cap04,cap17} to model the stellar continua. The emission lines are measured by fitting Gaussians to the observed data after subtracting the best-fit continuum models. In this paper, I use the Gaussian fitted emission line flux and equivalent width (\texttt{gflux} and \texttt{gew}) and absorption line indices in each pixels from the DAP. 

\subsection{The post-starburst galaxy sample}

The targets of interest are galaxies that would be classified as post-starburst galaxies if being placed at high-redshifts and observed under typical conditions then identified by common selection criteria. Therefore, the selection should be based on integrated properties rather than spatially resolved properties. 

In this paper, I adopt the selection based on spectral features. While photometric selections are now prevalent in high-redshift studies \citep[e.g.][]{whi12}, spectroscopical features are potentially more sensitive to the rapid decline of star-formation rates when high-S/N spectra are available \citep{wu20}. 

%As more deep spectroscopic surveys are coming \citep[e.g., StePS, MOONRISE,][]{cos19,mai20}, spectroscopic selection will be more common in the future. 

\subsubsection{Integrated spectra}
For each galaxy, I first collapse the data cube along the spatial axes, producing an 1D spectrum integrated over the entire IFU coverage. 
The values of spectral elements with \texttt{MANGA\_DRP3PIXMASK} that equal \texttt{DONOTUSE} are replaced by values interpolated from neiboring pixels. I apply no S/N cut on spaxels, i.e., even spaxels that sample the empty sky contribute to the integrated spectrum. This choice is closer to high-z observations, where the aperture of the spectrograph is a fixed angular size and can be larger than the target size. In practice, applying a low minimum S/N cut does not change the results. The low S/N spaxels contribute mainly random noises. I repeat the analysis using minimum S/N cuts up to 3 and find no difference relevant to this study.    

I use pPXF to separate the Balmers emission line and the underlying stellar absoprtion. Each 1D spectrum is fit by a combination of two sets of templates representing the gas emission and the stellar continuum. All emission lines are assumed to have Gaussian line shape with the same velocity and the velocity dispersion, but the strength of each line is a free parameter. The stellar template is a linear, optimal nonegative combiation of \citep{vaz99} SSP models with the Medium resolution INT Library of Empirical Spectra \citep[MILES,][]{san06} and \citet{gir00} isochrones. I adopt the definition of the H$\delta_a$ index in \citet{wor97} as my \ewhd. \edit1{The pseudo-continuum is defined at 4041.6\AA--4079.75\AA\ and 4128.5\AA--4161\AA, while the feature band is at 4083.5\AA--4122.25\AA.} The equivalent width of H$\alpha$ emission is the ratio between the flux of the best-fit H$\alpha$ emission model and the pseudo-continuum level at 6483\AA--6513\AA\ and 6623\AA--6653\AA\ as in DR15 \citep{wes19}. In this paper, I adopt the convention that absorption has positive equivalent width and emission is negative. 

\subsubsection{Selecting post-starburst galaxies}

\begin{figure}
	\centering
	\includegraphics[width=0.95\columnwidth]{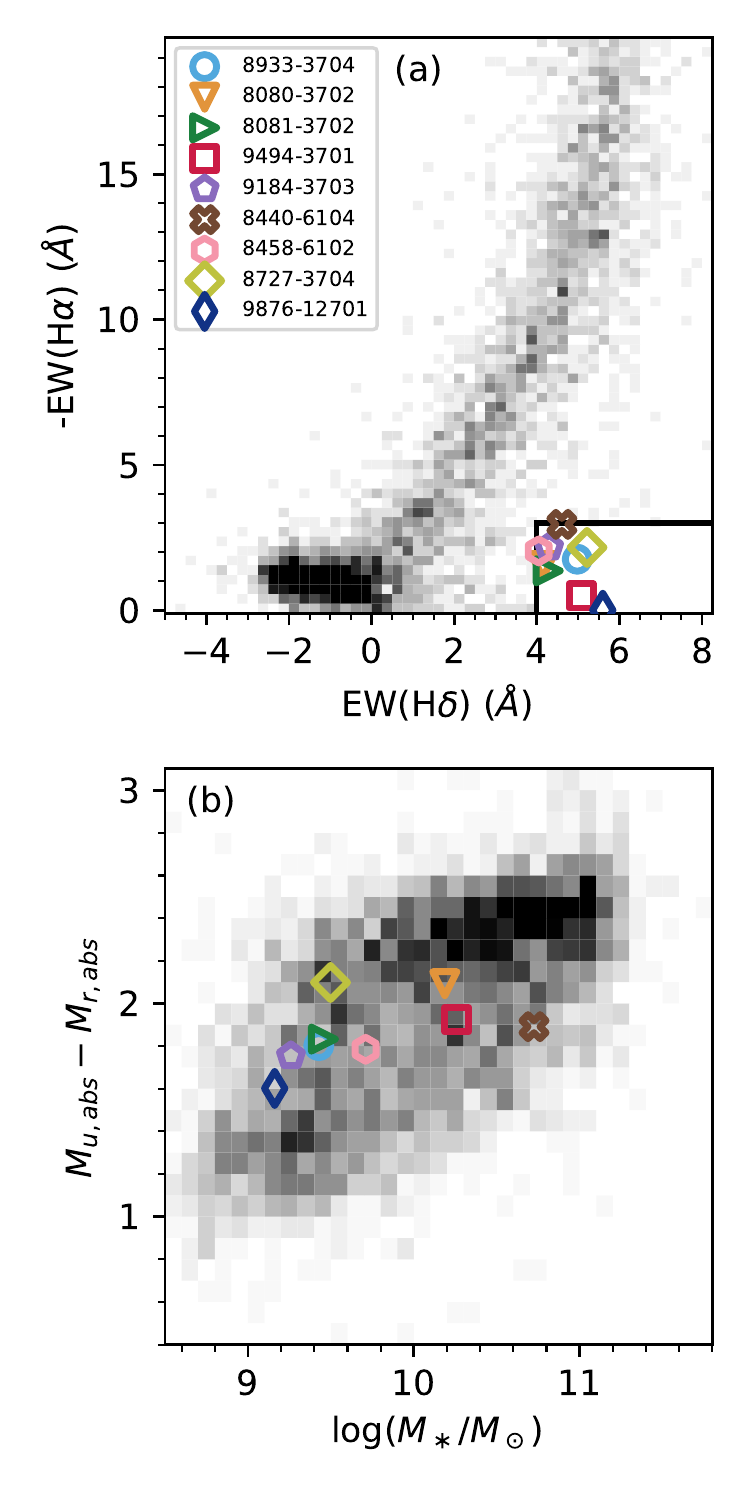}
	\caption{(a) The selection criteria of post-starburst galaxies. Gray 2D histogram is the distribution of \ewha\ and \ewhd\ of galaxies in the MaNGA DR15. The general population is located on a clear sequence. Post-starburst galaxies are outliers with strong H$\delta$ absorption and weak H$\alpha$ emission. (b) The color-stellar-mass diagram. Gray histogram is the MaNGA DR15 sample The colors and stellar masses are from the Nasa Sloan Atlas (NSA). Post-starburst galaxies are located between the red sequnce and the blue cloud. \label{fig:select} }
\end{figure}

\begin{figure*}
	\centering
	\includegraphics[width=0.95\textwidth]{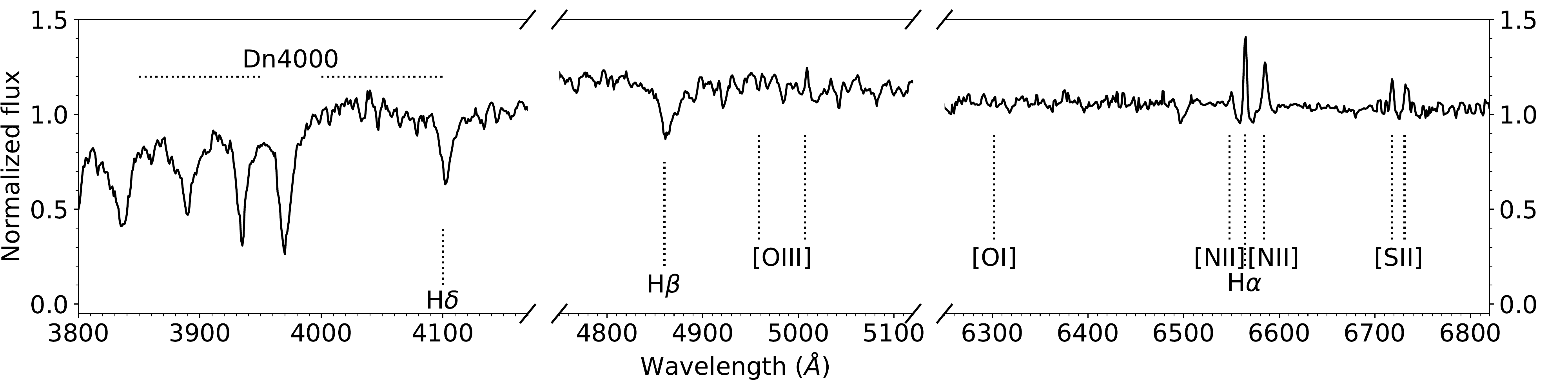}
	\caption{The stack spectra of 9 post-starburst galaxies. I only show the wavelength ranges that cover spectral lines relavant to this paper. The strong H$\delta$ absorption and the weak H$\alpha$ emission are by construction. Post-starburst galaxies also show weak [N\,II]$\lambda$6548,6584 and [S\,II]$\lambda$6717,6731 emission. On the contrary, H$\beta$, [OIII]$\lambda$4959,5007, and [OI]$\lambda$6300 do not show up clearly in the stacked 1D spectra. \label{fig:stack}}
\end{figure*}

I select post-starburst galaxies by H$\alpha$ emission and H$\delta$ absorption, a common combination used for low-$z$ galaxies. At higher redshifts, the H$\alpha$ emission is often substituted by the [O\,II]$\lambda$3727,3729 doublet because H$\alpha$ redshifts out of the otpical window. However, post-starburst galaxies often exhibit relatively high [O\,II]/H$\alpha$ ratios comaring to star-forming galaxies. Adopting the [O\,II] emission as the star-formation indicator leads to an incomplete and potentially biased sample \citep{yan06,lem10,lem17}. I thus adopt the Balmer emission line as the star-formation indicator. 

Fig.~\ref{fig:select}a shows the distribution of \ewha\ and \ewhd\ of the MaNGA sample (shaded area). Overall, the \ewha\ and \ewhd\ distribute on a clear sequence. Galaxies with stronger H$\alpha$ emission also have stronger H$\delta$ absorption. Galaxies with low star-formation rates occupy the lower left corner, mainly with $EW(H\delta)\lesssim 1$\AA\ and $EW(H\alpha)_{em} \gtrsim -3$\AA. Post-starburst galaxies should be located at the lower right corner: weak H$\alpha$ emission and strong H$\delta$ absorption. The fiducial criteria to select post-starburst galaxies from the 1D spectra are:
\begin{equation} \label{eq:select}
\begin{aligned}
\mbox{EW}(\mbox{H}\alpha)_{em} & \geq -3\mbox{\AA} \\
\mbox{EW}(\mbox{H}\delta) & \geq 4\mbox{\AA}.
\end{aligned}
\end{equation}

The boundaries on \ewha\ and \ewhd\ are somewhat subjective and a variety of combinations have been adopted in the literature. Samples selected with different boundaries have their own contamination and incompleteness and are also possible in different evolutionary stages \citep[e.g.,][]{yes14}. Investigating the effects of different cuts is beyond the scope of this paper. 

I further require that the velocity dispersions of best-fit templates of the gas and the stars should be smaller than 400 km s$^{-1}$ as a quality control of the pPXF fit. Such a high velocity dispersion is not expected for normal galaxies and likely indicates the failure of the fitting process. This is then confirmed by visual inspection on galaxies passing the EW cuts but rejected based on high velocity dispersion. 

Nine galaxies fulfill the criteria (Fig.~\ref{fig:select}a). The stellar masses range from $1.5\times 10^9 M_\odot$ to $5\times10^{10} M_\odot$ (Table.~\ref{tab:sample}). All post-starburst galaxies have $u-r$ colors between the red sequence and the blue cloud (Fig.~\ref{fig:select}b), based on the colors and stellar masses from the NASA Sloan Atlas. \footnote{\url{http://nsatlas.org/}} 

Figure~\ref{fig:stack} shows the median stack spectrum made from the 9 post-starburst galaxies. Each spectrum is first de-redshifted and resampled to the same wavelength grid. The flux is then normalized by the median flux between 4000\AA\ and 4100\AA\ before stacking. I only show wavelength ranges that used in this paper. By construction, the stack spectrum shows strong H$\delta$, as well as H$\beta$ absorption. In addition to H$\alpha$ emission, [N\,II]$\lambda$6548,6584 and [S\,II]$\lambda$6717,6731 also present in post-starburst galaxies. On the other hand, H$\beta$, [OIII]$\lambda$4959,5007, and [OI]$\lambda$6300 are not obvious in the stack 1D spectra. I will show that these diagnostic lines present in individual data cubes and carry important information (Section~\ref{sec:result} and Section~\ref{sec:dis})

\begin{deluxetable*}{ccrrccccchccc}
	%\rotate
	\tablecaption{Sample of post-starburst galaxies \label{tab:sample}}
	\tablehead{
		MaNGA ID & plateifu & RA & Dec & $z$ & $\log(M_\ast/M_\odot)$ & $R_e$ & b/a & P.A. & $C_{fiber}$ & EW(H$\alpha$)$_{em}$ & EW(H$\delta$) & D$_n$4000 \\
	        &          &  (deg)  &  (deg)   &          &                        & (arcsec) &  & (degree)    &             &  (\AA)      &  (\AA)         &           
}
	\startdata
1-457130 & 8933-3704 & 195.3305 & 27.8605 & 0.027 & 9.43 & 3.53 & 0.94 & 127.2 & 0.19 & -1.82 & 4.98 & 1.352 \\ 
1-38062 & 8080-3702 & 49.2289 & -0.0420 & 0.023 & 10.19 & 4.58 & 0.65 & 141.5 & 0.24 & -1.67 & 4.15 & 1.514 \\ 
1-38166 & 8081-3702 & 49.9469 & 0.6238 & 0.025 & 9.45 & 4.27 & 0.98 & 176.6 & 0.17 & -1.42 & 4.28 & 1.441 \\ 
1-384400 & 9494-3701 & 126.7557 & 21.7068 & 0.015 & 10.26 & 4.37 & 0.39 & 141.3 & 0.23 & -0.63 & 5.07 & 1.486 \\ 
1-153247 & 9184-3703 & 119.3653 & 33.2580 & 0.017 & 9.26 & 3.93 & 0.62 & 165.3 & 0.20 & -2.27 & 4.30 & 1.458 \\ 
1-26976 & 8440-6104 & 135.7590 & 40.4340 & 0.029 & 10.72 & 5.48 & 0.98 & 166.5 & 0.17 & -2.97 & 4.61 & 1.439 \\ 
1-606105 & 8458-6102 & 147.6643 & 44.3312 & 0.015 & 9.71 & 6.34 & 0.61 & 23.7 & 0.10 & -2.11 & 4.05 & 1.468 \\ 
1-51631 & 8727-3704 & 56.3971 & -7.3843 & 0.028 & 9.50 & 4.62 & 0.63 & 168.4 & 0.16 & -2.18 & 5.21 & 1.460 \\ 
1-456850 & 9876-12701 & 194.6345 & 28.3780 & 0.020 & 9.16 & 12.51 & 0.73 & 75.9 & 0.05 & -0.00 & 5.60 & 1.494 \\ 
	\enddata
\end{deluxetable*}

\section{Resolving post-starburst galaxies}
\label{sec:result}

I will present 4 features that are frequently disucssed; (1) The locations of post-starburst regions, (2) the gradient of absorption line indices, which are often used to infer the age gradient, (3) the emission lines and their origins, and (4) the kinematics. 

\subsection{Spatial distribution of post-starburst regions}

%*** Redo the image plot. Delete one galaxy and modify the color label. ***
\begin{figure*}
	\centering
	\includegraphics[width=0.95\textwidth]{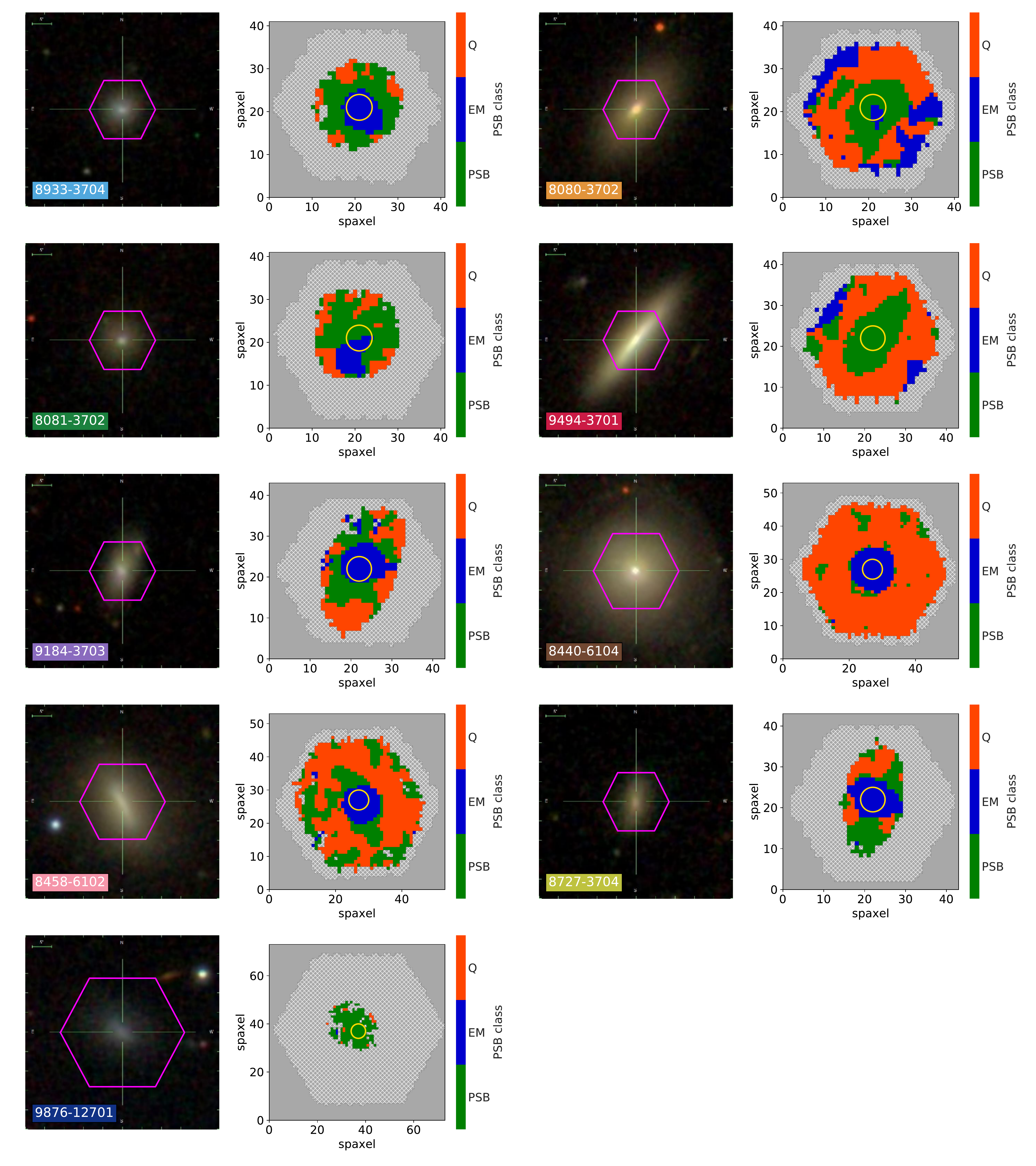}
	\caption{The pseudo-color images and the spectral types of spaxels of post-starburst galaxies. The magenta hexagons are the apertures of MaNGA IFU. For the spectral classification, only spaxels of $S/N \geq 3$ are used. Red, green, and blue areas are quiescent, post-starburst, and emission regions, respectively. The yellow circles correspond to the apertures of the Sloan fiber. Post-starburst galaxies are inhomogeneous and a large fraction of area is occupied by quiescent and emission spaxels. Emission-line centers are prevelent among post-starburst galaaxies.}
	\label{fig:psb_class}
\end{figure*}

Galaxies are not homogeneous systems. To present the salient features of spectral properties within post-starburst galaxies, I classify spaxels into 3 types:

\begin{equation} \label{eq:class}
\begin{aligned}
\mbox{Post-starburst: }& EW(H\alpha)_{em} \geq -3\AA \\
&  \&\ EW(H\delta) \geq 4\AA \\
\mbox{Quiescent: }& EW(H\alpha)_{em} \geq -3\AA \\
&  \&\ EW(H\delta) < 4\AA \\
\mbox{Emission: }& EW(H\alpha)_{em} < -3\AA 
\end{aligned}
\end{equation}

This classification follows the definition of the sample selection: setting the demacartion of emission and absoprtion strengths at $\mbox{EW(H}\alpha\mbox{)}_{em} = -3$\AA\ and $\mbox{EW(H}\delta\mbox{)} = 4$\AA, respectively. The criteria for post-starburst spaxels is the same as selecting post-starburst galaxies. Quiescent spaxels are regions void of on-going star-formation and with older stellar population than post-starburst spaxels. All other spaxels, those with H$\alpha$ emission, are emission spaxels, regardless of the H$\delta$ absorption strength. 

Fig.~\ref{fig:psb_class} shows the color images and the spatial distributions of each class of spaxels. Firstly, post-starburst spaxels only consist of half or even smaller fraction of area that is observed. An extreme example is 8440-6104, only $<10\%$ of the area is classified as post-starburst. However, I will show in the next section that these `quiescent' regions at the outskirts still have relative strong H$\delta$ absorption.

Another noticable feature is the prevalent central emission-line regions. The centers of 5 galaxies (8933-3704, 9184-3703, 8440-6104, 8458-6102, and 8727-3704) are occupied by emission spaxels. The yellow circles in Figure~\ref{fig:psb_class} shows the apertures of the Sloan fiber. These 5 galaxies would be classified as emission-line galaixes based on Sloan fiber spectra. I will further discuss the properties of emission lines in Section~\ref{sec:ha}. 

This simple inspection (1) demonstrates that post-starburst galaxies selected from spatially-unresolved data are instrinsically inhomogeneous and (2) warns a direct comparison among samples selected using different aperture sizes, e.g., spectroscopic samples at different redshifts. 

\subsection{Spectral indices}
\begin{figure} 
	\centering
	\includegraphics[width=0.95\columnwidth]{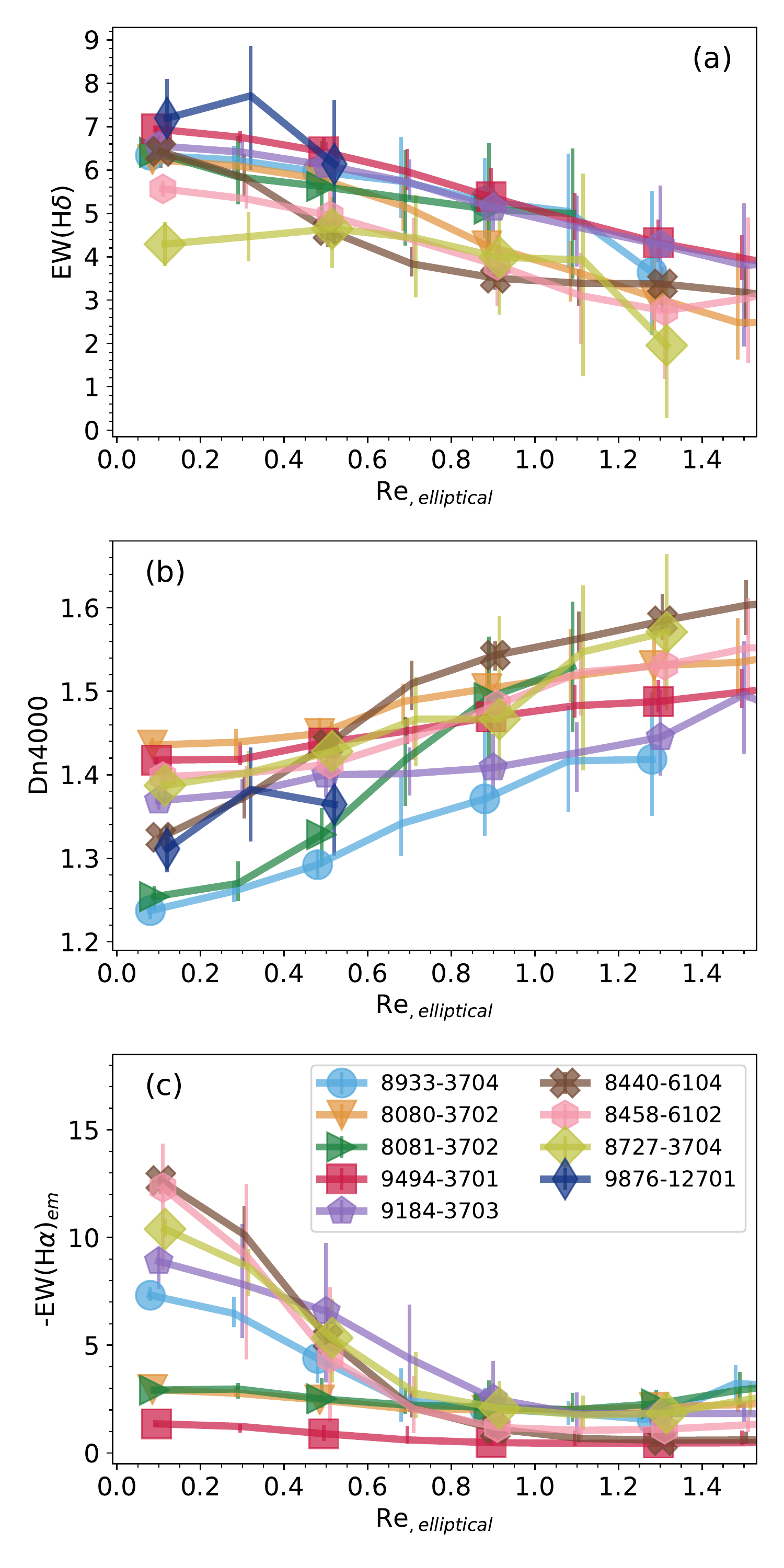}
	\caption{The spectral indices as a function of elliptical effective radii. Each datapoint is the median indice at a given radius with a bin width of 0.2 Re. The data is presented out to 1.5 Re or the outmost bin where at least one spaxel with $S/N \geq 3$ in the bin. \edit1{The formal measurement uncertainties after binning is in general very small even with the covariance taken into account. The errorbars plotted here are either the median uncertainties of each spaxel in the bin or the 16th or 84th percentiles of the index distribution in the bin. The larger of the two cases is plotted.} (a) \ewhd. (b) \dn. Most post-starburst galaxies have larger \ewhd\ and smaller \dn\ at the centers. (c) \ewha. The H$\alpha$ emission in post-starburst galaxies tend to be located at the centers. The galaxy 9876-12701 has no high S/N H$\alpha$ detection thus is not shown in panel (c). Markers are plotted every 2 bins \edit1{and small horizontal shifts are applied} for clarity. \label{fig:ind_r}}
\end{figure}

I plot \ewhd, \dn, and \ewha\ as a function of \edit1{elliptical radius} (\texttt{ellcoo\_r\_re} in MaNGA maps, hereafter $R_{e,ell}$). \edit1{The $R_{e,ell}$ is the NSA elliptical Petrosian effective radius in the r-band. The position angles and axis ratios of the elliptical apertures are listed in Table~\ref{tab:sample}.} The data points are the median indices in bins of every 0.2$R_{e,ell}$ (Fig.~\ref{fig:ind_r}). Overall, post-starburst galaxies have stronger H$\delta$ absorption and smaller \dn\ in the center. I measure the indice gradients of \ewhd\ and \dn\ from galaxy center to $1.5 R_e$ or outmost available data points. All post-starburst galaxies have negative \ewhd\ and positive \dn\ gradients, with medians of $\Delta\mbox{EW(H}\delta\mbox{)}/\Delta R_e = -2.0$\AA\ and $\Delta D_n4000/\Delta R_e = 0.13$. These gradients are distinct from the average values measured from the MaNGA survey, which are almost flat or even with opposite signs \citep{wan17} and suggest that the centers of post-starburst galaxies have younger stellar ages as can be intuitively inferred from Fig.~\ref{fig:psb_class}; the quiescent spaxels are located mostly at outskirts, the post-starburst spaxels distribute near or at the centers, and the emission-line spaxels are mainly found at the centers.  

%These gradients are distinct from the average values measured from the MaNGA survey, which are almost flat or even with opposite signs \citep{wan17}. I construct a mass-color matched sample from MaNGA to compare with post-starburst galaxies. For each post-starburst galaxy, I randomly select 5 galaxies with stellar masses ($\pm0.2$~dex) and restframe $u-r$ color ($\pm0.1$) for comparison. The shaded areas in Fig.~\ref{fig:ind_r} are the distributions (16th, 50th, and 84th percentiles) of mass-color-matched sample. The comparison sample has much flatter gradients ($\Delta\mbox{EW(H}\delta\mbox{)}/\Delta R_e = 0.6$\AA\ and $\Delta D_n4000/\Delta R_e = 0.01$) and overall smaller \ewhd\ and larger \dn. Even the two samples have similar stellar mass and color distributions, the intrinsic distributions of stellar populations are drastically different.

It is worth noting that even at the outskirts, the stellar populations of post-starburst galaxies are still relatively young. Most post-starburst galaxies have $EW(H\delta) \geq 3\AA$ at 1.5 $R_{e,ell}$, which belong to the `young' population like star-forming galaxies \citep{kau03b}. On the other hand, the low \ewha\ at the outskirts indicates low star formation rates (Fig.~\ref{fig:ind_r}c). The combination suggests that the star-formation activities also had declined rapidly at outskirts of post-starburst galaxies. 

\subsection{H$\alpha$ emission}
\label{sec:ha}

\begin{figure}
	\centering
	\includegraphics[width=0.95\columnwidth]{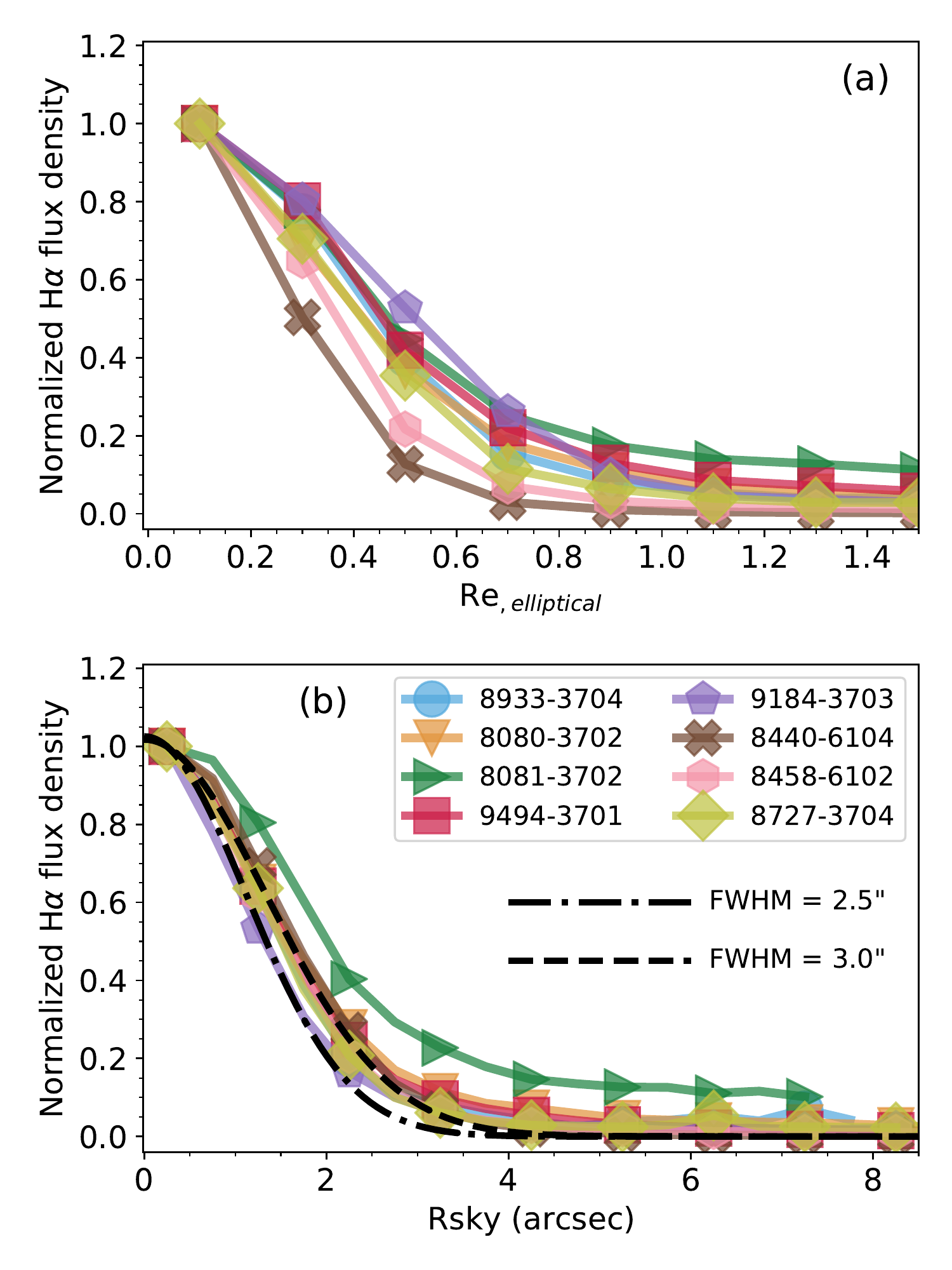}
	\caption{The radial dependence of H$\alpha$ fluxes of post-starburst galaxies. (a) The normalized surface flux density as a function of elliptical effective radii. I first calculate the median flux densities from spexals at similar elliptical radii, using bins of 0.2$R_e$. The median fluxes are then normalized by the values of the innermost bin. The H$\alpha$ emission in post-starburst galaxies concentrate in galaxy centers. (b) The normalized surface flux densities in circular rings of every 0.5". Markers are plotted every 2 bins for clearity. The dashed-dotted line and the dashed line are Gaussians with FWHM of 2.5" and 3.0" arcsec, respectively. The H$\alpha$ flux densities of most post-starburst galaxies are well approximated by Gaussians of FWHM smaller than 3.0", only slightly larger than the MaNGA PSF of $\sim$2.5". The intrinsic sizes of the emission sources should thus be $\lesssim 1$~kpc. \label{fig:ha_r}}
\end{figure}

Contrary to the absorption indices, the \ewha\ profiles show some variations amongst the post-starburst sample (Fig.~\ref{fig:ind_r}c). Firstly, MaNGA 9876-12701 (not shown in the figure) has no spaxel with detectable H$\alpha$ emission ($S/N \geq 3$). For the other 8 post-starburst galaxies, 3 of them have weak ($\mbox{EW(H}\alpha\mbox{)}_{em} \lesssim 3$\AA) and nearly constant $\mbox{EW(H}\alpha\mbox{)}_{em}$ profiles out to $1.5 R_{e,ell}$. The remaining 5 post-starburst galaxies have \ewha\ peaks at the centers, as high as $\sim12$\AA, then sharply drops to $\mbox{EW(H}\alpha\mbox{)} \lesssim 3$\AA\ beyond $R_{e,ell}$. %The comparison sample (the shaded area), on the other hand, have overall flatter profiles and larger \ewha\ at outskirts. 

Instead of the equivalent width, Fig.~\ref{fig:ha_r}a shows the flux density profile of each post-starburst galaxy after normailized by the value at the center. The H$\alpha$ emission in post-starburst galaxies concentrates in the centers. At $1R_{e,ell}$, the flux density is in generally only $\sim10\%$ of the peak. %On the other hand, the flux density profiles of the comparidson sample (shaded area) are more extended. 

Fig.~\ref{fig:ha_r}b shows the normalize flux profiles as a function of angular distances from galaxy centers, i.e., not taking into account the $R_{e,ell}$ and inclinations. The similarity is stricking. Six out of 8 galaxies appear nearly identical to each other in the inner $\sim2"$ and all drop to below 10\% of the peak fluxes beyond $\sim4"$. One galaxy (9184-3703) is slightly broader and one galaxy (9184-3703) is slighly narrower, while other 6 galaxies appear nearly identical to each other in the inner $\sim2"$ and all drop to below 10\% of the peak fluxes beyond $\sim4"$.

The dash-dotted and the dashed lines are Gaussians with FWHM of 2.5" and 3.0", respectively. The typical PSF of MaNGA at r-band (\texttt{rFWHM}) is $\sim$2.5". The normalized flux profiles of most post-starburst galaxies are only slightly broader than the PSF, which suggests that the central emission is compact. Under a simplified assumption that the central emitting sources also have 2D Gaussian profiles projected on the sky, the observed FWHM suggests that the sources have intrinsic FWHM$\simeq1.6"$, corresponding to $\sim0.9$~kpc at the median redshift ($z=0.025$) of the sample.

\subsection{Emission v.s. absorption}

\begin{figure}
	\centering
	\includegraphics[width=0.95\columnwidth]{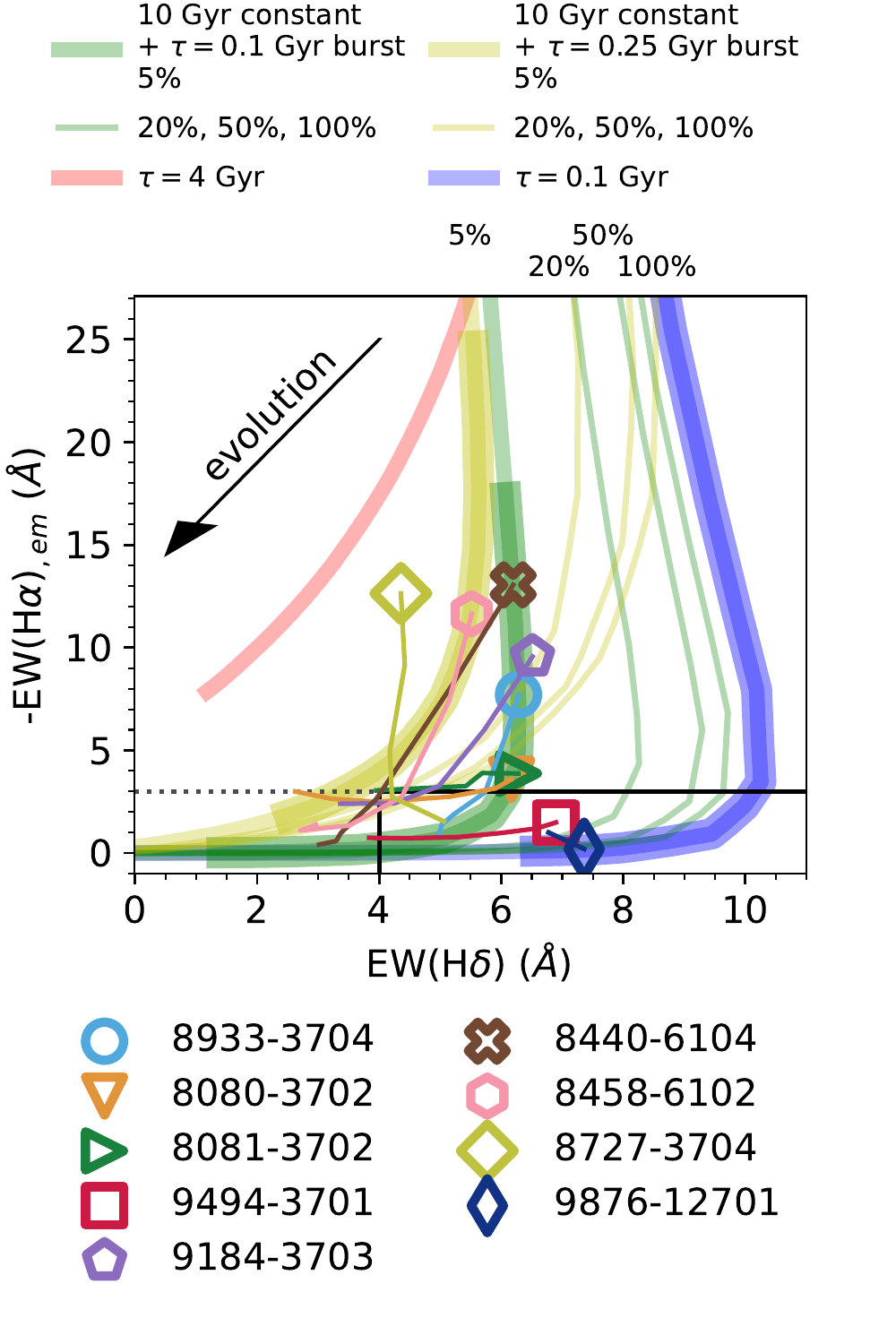}
	\caption{The \ewha\ and \ewhd\ of post-starburst galaxies at different radii. Each galaxy is represented by a line. The centers are marked by the symbol and the other ends of the lines are the outmost radii that both \ewha\ and \ewhd\ can be measrued. The color strips are model evolutionary tracks. The red and blue stripes are exponentially-decay models with e-folding times of 4 and 0.1~Gyrs, respectively. The green and yellow strips are galaxies form with a constant star-formation for 10~Gyr then an extra 5\% exponential burst with e-folding times of 0.1 and 0.25~Gyrs, repectively. The thick segements highlight 0.2 to 1.0~Gyr after the onset of the exponential decay. \edit1{The green and yellow thin lines are also the constant-then-burst formation histories, with extra 20\%, 50\%, and 100\% bursts.} Most post-starburst galaxies are located closer to models with old stellar populations plus weak recent burst and truncation. The outskirts are overall older, $\sim1$~Gyr after the onset of the burst. \label{fig:hahd}}
\end{figure}

The H$\alpha$ emission and H$\delta$ absorption are sensitive to different time scales of star-formation activities. Galaxies with different star-formation histories evolves along different trajectories on the \ewha---\ewhd\ plane. 

Combining Fig.~\ref{fig:ind_r}a and Fig.~\ref{fig:ind_r}c, Fig.~\ref{fig:hahd} shows each post-starburst galaxies on the \ewha---\ewhd\ plane in bins of $R_{e,ell}$. Each galaxy is represented by a line, where the center is labeled by the marker and the outskirt is the other end of the line. 
The thick strips are model evolutionary tracks with different star-formation histories for illustration. The red and blue lines are exponentially-decay models: $\psi(t) \propto \exp(-t/\tau)$, where $\tau$ are 4 and 0.1~Gyr, respectively. The green and yellow strips are galaxies form with a constant star-formation rate for 10~Gyr then followed by an exponential burst adding 5\% of stars with $\tau$ of 0.1 and 0.25~Gyrs, respectively. These two models represent galaxies with long formation time, as expected for local galaxies, then just shut down rapidly recently. The thick segments are 0.2 to 1.0~Gyr after the onset of the expoention decay. \edit1{Other green and yellow thin lines are also the constant-then-burst star-fomation histories but with 20\%, 50\%, and 100\% bursts.} All models are based on the \citet{che16} update of the \citet{bc03} model, with a dust attenuation of $A_V=2$ for young ($<10$~Myr) stars and $A_V=1$ for the rest and the \citet{cal00} extinction curve. 

While the exact evolutionary tracks are affected by multiple factors \edit1{whose effects degenerate on the 2-dimensional plane} and there is no guarantee that the centers and the outskirts should have gone through the same star-formation histories, the simple models should still illustrate the salient points of the histories of these galaxies. 
Broadly speaking, the loci of most post-starburst galaxies are close to models with the long star-formation preiod, as expected for local galaxies, and a rapid decline of star-formation rate. A pure short-$\tau$ model (the blue line) produces galaxies with too strong H$\delta$ absorption. In addition, we do not expect such formation histories for local galaxies with such stellar masses. On the contrary, A pure long-$\tau$ model (the red line) produces galaxies with too strong H$\alpha$ emission: the star formation is still active when the overall stellar population is already aged. 

\edit1{The \ewha\ and \ewhd\ of post-starburst galaxies indicates that small bursts (5\%) can produce the indices across the entire galaxies, while stronger bursts event are also possible.}
Although the outskirts are not classified as `post-starburst' based on the fiducial classification scheme, their locations on the \ewha---\ewhd\ plane suggest that the star-formation rates also had declined rapidly. They are just at the later stages of the evolution, $\lesssim 1$~Gyr after the onset of the quenching process; therefore, the stellar populations are older and exhibit weaker H$\delta$ absoprtion. On the other hand, the centers are at earlier evolutionary stages where the residual star-formation activities are still detectable. 

\subsection{Emission line properties}
\label{sec:bpt}
\begin{figure*}
	\centering
	\includegraphics[width=0.95\textwidth]{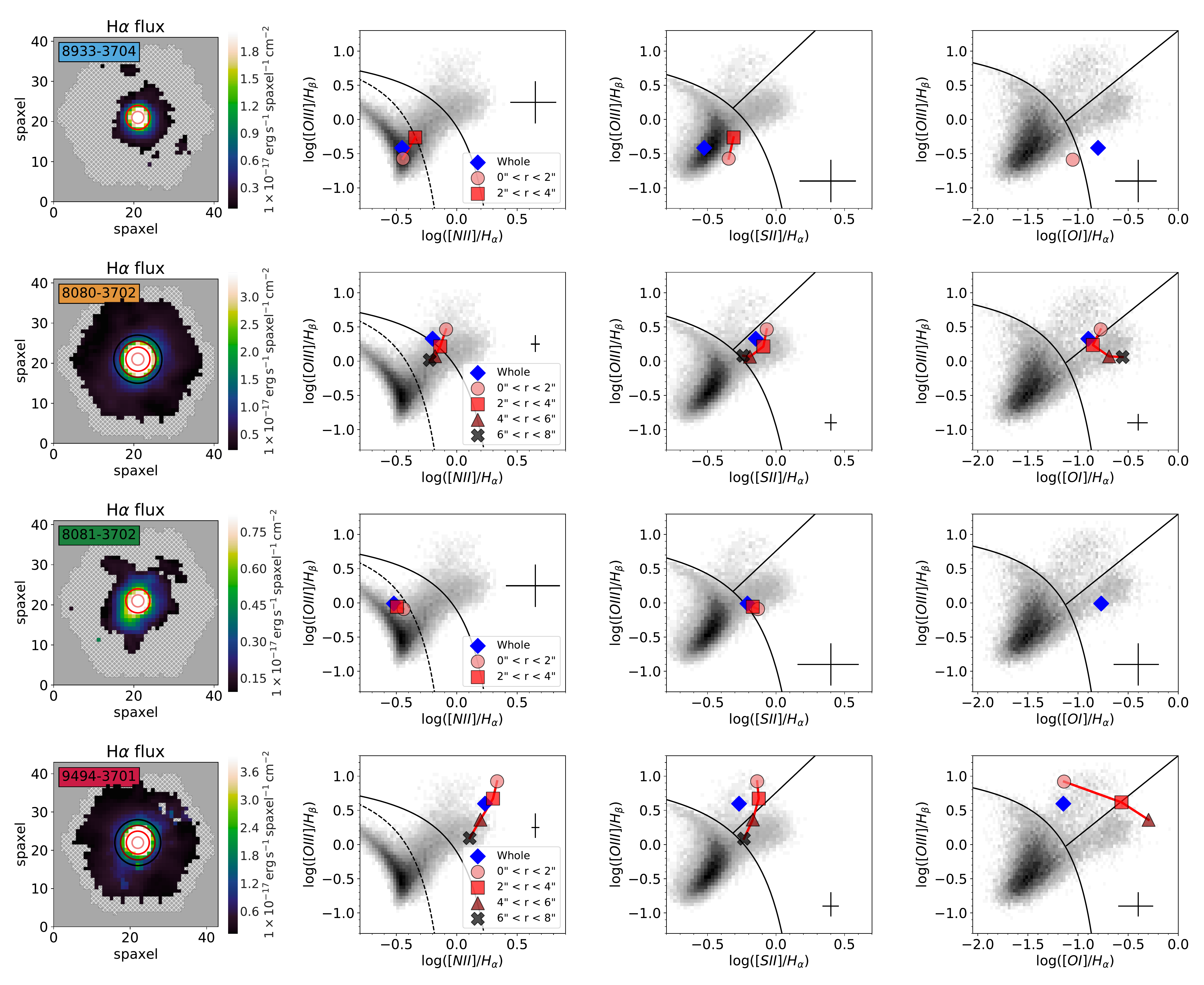}
	\caption{The diagnostic emission line ratios of post-starburst galaxies at different radii. Each row presents one galaxy. The first panel shows the map of H$\alpha$ flux. The following three panels show the diagnostic line ratios: [N\,II]/H$\alpha$, [S\,II]/H$\alpha$, [O\,I]/H$\alpha$, and [O\,III]/H$\beta$ at given radii, calculated from the integrated fluxes in rings of every 2" from the galaxy centers. The circles (pink), squares (red), triangles (brown), and crosses (black) represent the line ratios from the center to the outskirts. The diamonds (blue) are the line ratios calculated from summing up spaxels at all radii. In all panels, only flux measurements with $S/N>3$ are used for calculation and plotting. Galaxies with no high-S/N spaxels to calculate the line ratios are not plotted here. For all galaxies, the line ratios at outskirts are located at the LINER/composite region, similar to elliptical galaxies. On the contrary, the line ratios at the centers vary. Two galaxies (9494-3701 and 8080-3702) have AGN-like line ratios in the center, while other 5 galaxies have star-forming line ratios at the centers. One galaxy (8081-3702) does not show clear gradient. The radial dependence of the emission line ratios indicates a variety of central ionizing sources in post-starburst galaxies. \label{fig:bpt}}
\end{figure*}

\begin{figure*}
	\centering
	\figurenum{7 cont.}
	\includegraphics[width=0.95\textwidth]{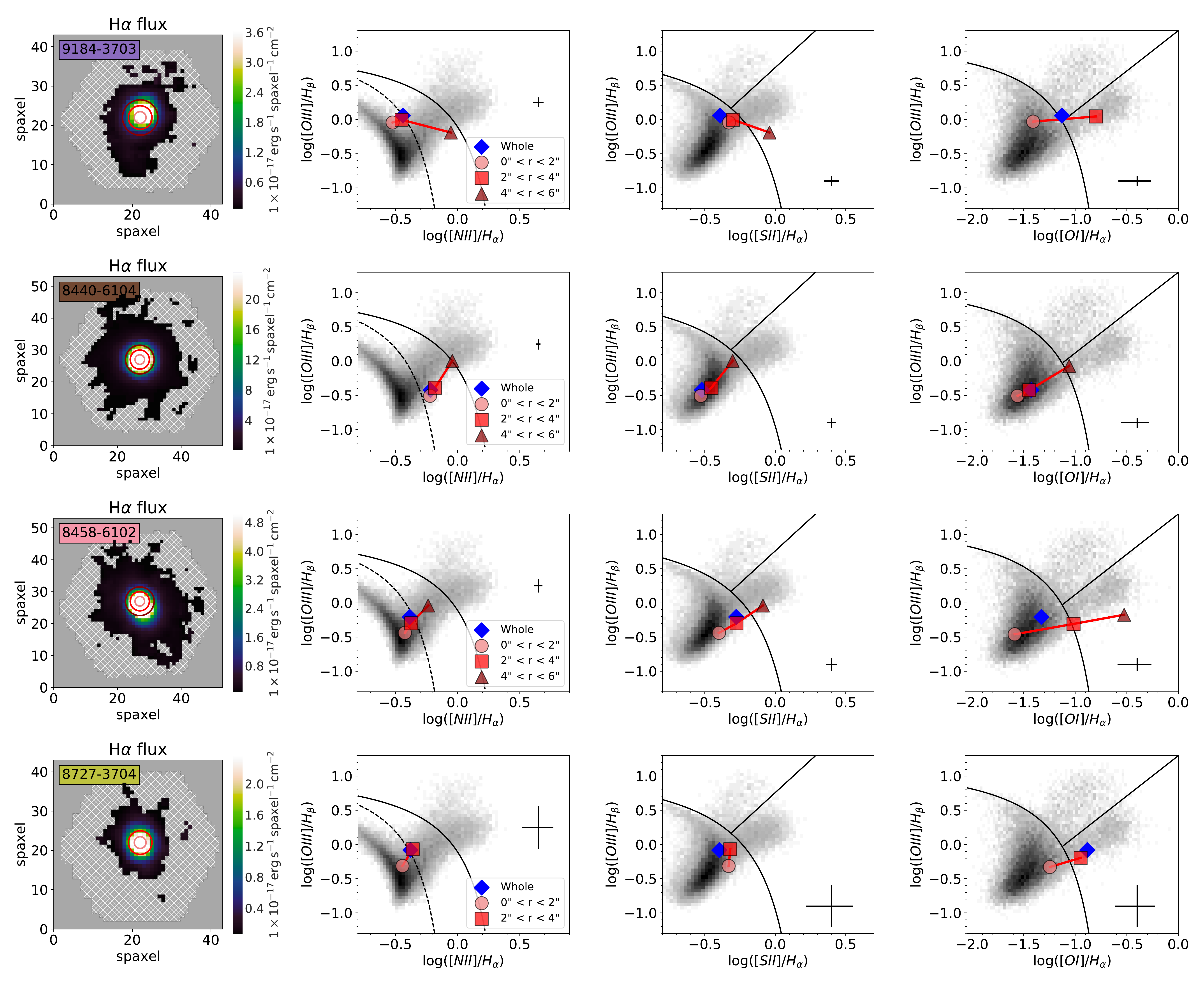}
	\caption{\label{fig:bpt2}}
\end{figure*}

\figsetend

The integrated spectra of post-starburst galaxies exhibit a series of emission lines (Fig.~\ref{fig:stack}) that can be produced by a variety of mechanisms. The intensity ratios of emission lines provide clues for the ionizing sources \citep[so-called BPT diagrams,][]{bal81,vei87}.

\subsubsection{BPT diagrams from resovled measurements}
Fig.~\ref{fig:bpt} shows the diagnostic line ratios at different radii of each post-starburst galaxy. Each data point presents the median line ratios of spaxels in radial bins of 2". Measurements from different radial bins are plotted with different colors and shapes. The circles on the H$\alpha$ maps (the first column of Fig.~\ref{fig:bpt}) are the corresponding apertures. In all panels, only lines with $S/N\geq3$ are used for generating the plots. The blue diamonds are measurements from the integrated spectra. 

The line ratios change along with distances from galaxy centers. For most of the post-starburst galaxies, the emission from the central 2" and the outmost rings belong to different classes on the BPT diagrams, suggesting different ionizing sources at galaxy centers and outskirts. Furthermore, the emission line ratios at the centers also reveal the heterogeneous origins of the ionizing sources. Both AGN and star-forming regions present at the centers of post-starburst galaxies. 

The most outstanding case for AGN is 9494-3701. All line ratios indicate an AGN at the center.  Another galaxy with strong indication of AGN is 8080-3702, which has [N\,II]/H$\alpha$ and [O\,I]/H$\alpha$ consistent with AGN and [S\,II]/H$\alpha$ near the boundary of LINER and Seyfert. Both 8080-3702 and 9494-3701 have their center 3" filled with mainly post-starburst spaxels (Fig.~\ref{fig:psb_class}).

For the other 6 galaxies, the line ratios at the centers are almost all consistent with powering by star-forming activities (except 8081-3702) but often offset from the main star-forming branch (the background gray shaded area in Fig.~\ref{fig:bpt}) toward the LINER region. These are also galaxies with large H$\alpha$ equivalent width at centers (Fig.~\ref{fig:ind_r}c) and with their center 3" filled with mainly emission line spaxels (Fig.~\ref{fig:psb_class}). The larger H$\alpha$ equivalent widths and line ratios suggests that these post-starburst galaxies have low-level star-formation activities concentrated within the central $\lesssim 1$~kpc (Sec.~\ref{sec:ha}). 

Contrary to the heterogeneoity at the centers, the line ratios at outskirts of post-starburst galaxies show high similarities. 
The outmost points are all located on the AGN branch on the [N\,II]-H$\alpha$ diagram and either in or close to the `composite' region, indicating that the emission at outskirts of post-starburst galaxies do not originate from star-forming activities. Furthermore, the [S\,II]/H$\alpha$ and [O\,I]/H$\alpha$ are higher than typical star-forming galaxies and mostly located in the LINER region. 
The LINER-like line ratios, in addition to the low \ewha\ (Fig.~\ref{fig:ind_r}c), is similar to emission in quiescent galaxies or quiescent regions in IFU observations and often attributed to evolved stars \citep{cid10,sin13}. 

%mix of emission and post-starburst spaxels. 
%emission spaxels are off-center. 
%This is also the one galaxy which has a broader normalize H$\alpha$ flux. Examining the H$\alpha$ map and continuum flux map find that the peaks are offset by ? arcsec. 
%weak H$\alpha$
%with central \ewha$>-3$\AA. 

%*This offset may be suggesting that the emission is blended with LINER region. As the aperture move outwards, the data approach LINER region more \citep[also see][]{sin13}*

In short, the line emission of post-starburst galaxies consists of two distinct components. The first is an extended, LINER-like emission, which can be attributed to evolved stars. The second is a compact ($\lesssim1$~kpc) component at the centers of galaxies. This component can be powered by either AGN or low-level star-formation activities. 

%As the total flux is dominated by the central compact component (Fig.~\ref{fig:ha_r}), the line ratios of the integrated spectra are generally similar to those from the center 2" (Fig.~\ref{fig:bpt}). For the most frequently used pairs of lines, \niiha\ and \oiiihb, the differences are mostly $\lesssim 0.1$~dex and $\lesssim 0.2$~dex, respectively. These differences, while non-negligible, do not change the classification of the central ionizing sources in these 8 post-starburst galaxies using the \niiha-\oiiihb\ diagram. When the fluxes have low S/N, the uncertainties in classification comes mainly from the measurement rather than the aperture effect (e.g., 8081-3702 and 8727-3704).

\subsubsection{Gas metallicity}
\begin{figure}
	\includegraphics[width=0.95\columnwidth]{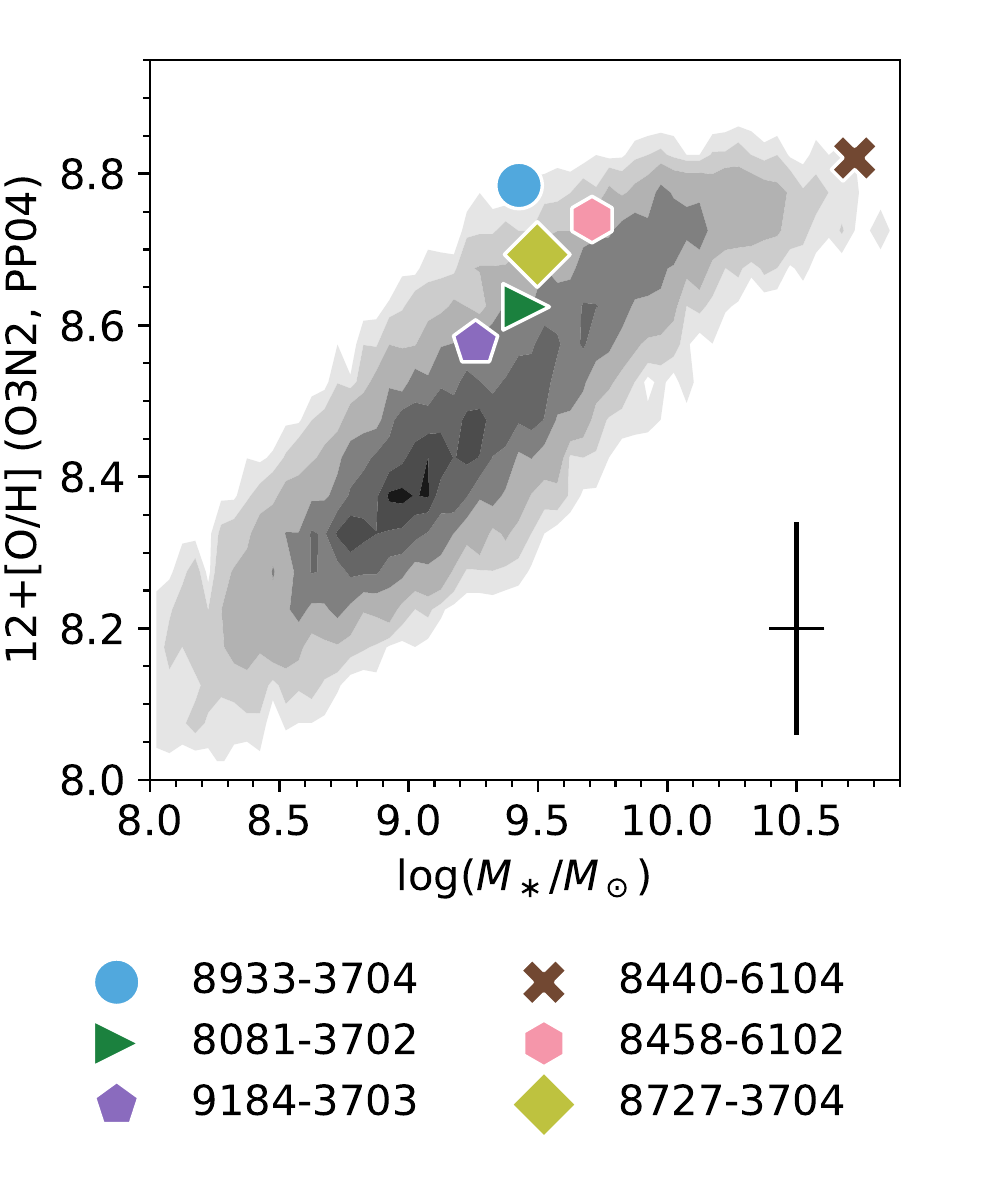}
	\caption{The gas-phase metallicity of post-starburst galaxies. The gray shaded area is distribution of metallicity of galaxies at similar redshift. The metallicities of post-starburst galaxies are higher than the average, consistent with the expectation based on their low star-formation rates. The errobar is the 0.14~dex systematic uncertainty from the calibration. The uncertainties from the measurements are all $<0.05$~dex. \label{fig:mz}}
\end{figure}

The metal content in the interstellar medium (ISM), the gas-phase metallicity, captures the histories of star-formation and gas flows in and out of galaxies. The gas-phase metallicity is often inferred by fluxes ratios of strong emission lines.

In principle, the line fluxes need to be corrected for dust attenuation, usually estimated by comparing the observed flux ratios of Balmer emission lines to theoretical intrinsic values. However, the H$\beta$ emission in some post-starburst galaxies are relatively weak, which introduces large uncertainties to the estimated dust attenuation thus the intrinsic fluxes. 

I thus adopt the empirical calibration of \citet{pp04} based on the O3N2 index:
\begin{equation}
	12 + \log(O/H) = 8.73 - 0.32 \times \mbox{O3N2}, 
\end{equation}
where O3N2 is defined as:
\begin{equation}
	\mbox{O3N2} \equiv \log \frac{[O\,III]5007/H\beta}{[N\,II]/H\alpha}.
\end{equation}
This index has the advantage that it uses the flux ratios of emission lines at similar wavelengths, thus minimizing the uncertainties introduced by attenuation correction. 

To estimate the gas-phase metallicity of the star-forming regions in post-starburst galaxies, I use the line fluxes at the central pixel of each data cube. The emission line regions appear to be located at the centers \edit1{and nearly unresolved by MaNGA Fig.~\ref{fig:ha_r}.} Therefore, the central pixel is expected to be less affected by other ionizing sources, e.g., the underlying older stellar populations dominating at the outskirst (Fig.~\ref{fig:bpt}). 

Figure~\ref{fig:mz} shows the gas-phase metallicities and stellar masses of post-starburst galaxies. Two galaxies with Seyfert-like emission (8080-3702 and 9494-3701, see Sec.~\ref{sec:bpt}) are not plotted. The emission lines in 9876-12701 are too weak for estimating the metallicity. I estimate the uncertainties in $12+\log(O/H)$ by repeating the measurement 1000 times, each time the fluxes are disturbed by values that drawn from a Gaussian distribution with $\sigma$ equal to the flux uncertainties. The uncertainties in $12+\log(O/H)$, determined from the 16th and the 84th percentiles of the 1000 measurements, are all $<0.05$~dex, much smaller than the uncertainties inherited from the calibration \citep[0.14~dex, the errorbar in Fig.~\ref{fig:mz},][]{pp04}.

I select star-forming galaxies in the NSA for comparison. Firstly, galaxies need to have redshifts within the range of the post-starburst galaxy sample, $0.015 < z < 0.029$. Then, the emission lines need to have $EW(H\alpha) < -3$\AA\ and star-formation-like \niiha\ and \oiiihb\ line ratios. Lastly, all 4 lines need to have $S/N \geq 3$. The gray filled contours in Fig.~\ref{fig:mz} are the distribution of the comparison sample on the mass-metallicity plane. 

Overall, the gas-phase metallicities at the centers of post-starburst galaxies are $\gtrsim0.1$~dex larger than the average star-forming galaxies of similar masses. I also estimated the metallicities of post-starburst galaxies using the fluxes in the Sloan fiber, the results differ by only $\lesssim0.02$~dex. 

\subsection{Half-light radii}
\begin{figure}
	\centering
	\includegraphics[width=0.95\columnwidth]{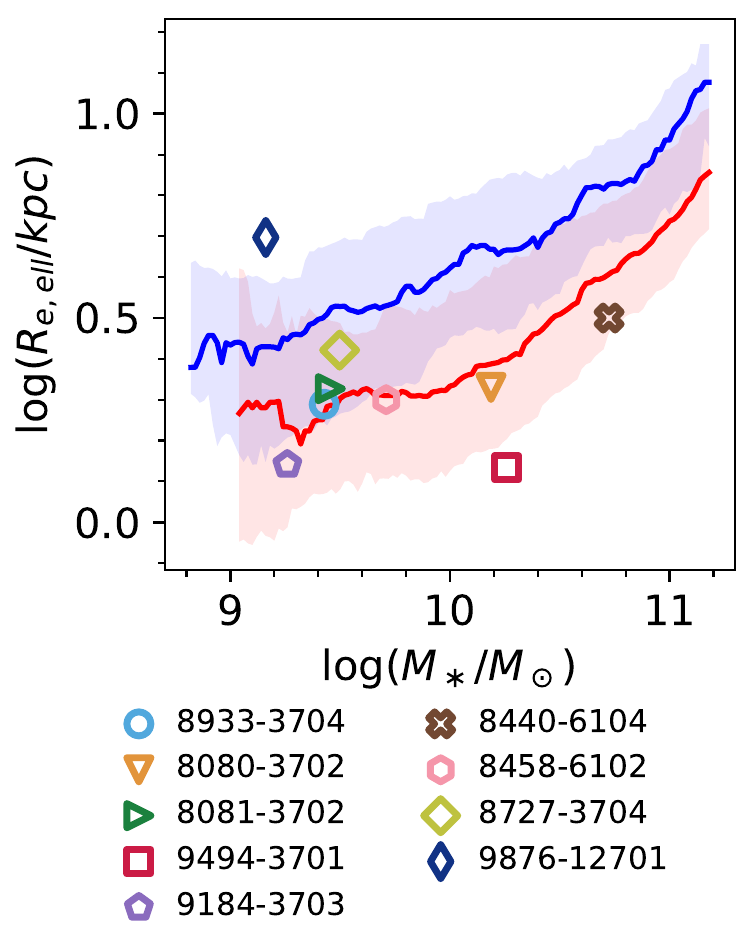}
	\caption{The size $R_{e,ell}$ and the stellar masses of post-starburst galaxies and the MaNGA sample. The red and blue solid lines are the median mass-size relations of red and blue galaxies in the MaNGA survey, respectively. Shaded areas represent the 16th and 84th percentiles at a given stellar mass. Post-starburst galaxies have on average similar sizes as red galaxies. \label{fig:MR}}
\end{figure}

\edit1{Fig.~\ref{fig:MR} plots the stellar masses and $R_{e,ell}$ of post-starburst galaxies and the 16th, 50th, and 84th percentiles of the sizes at fix stellar masses of the blue and red galaxies in the MaNGA survey. I take $M_{u,abs} - M_{r,abs} = 1.8$ as the fiducial demacartion between the red and blue populations (see Fig.~\ref{fig:select}b). }

\edit1{The $R_{e,ell}$ of most post-starburst galaxies are located below the 16th percentiles of blue galaxies and comparable to the median sizes of red galaxies. The light distribution of 9876-12701 is extremly extended; this galaxy is a peculiar object in all respects (see previosu sections). Galaxy 9494-3701 on the contrary has a small $R_{e,ell}$. This is a highly inclined galaxy with central emission-line region dominated by AGN (Fig.~\ref{fig:psb_class} and Fig.~\ref{fig:bpt}). I will discuss the implication of the sizes of post-starburst galaxies in Sec.~\ref{sec:dis}. }

\subsection{Stellar kinematics}
\label{sec:kin}

\begin{figure*}
	\centering
	\includegraphics[width=0.95\textwidth]{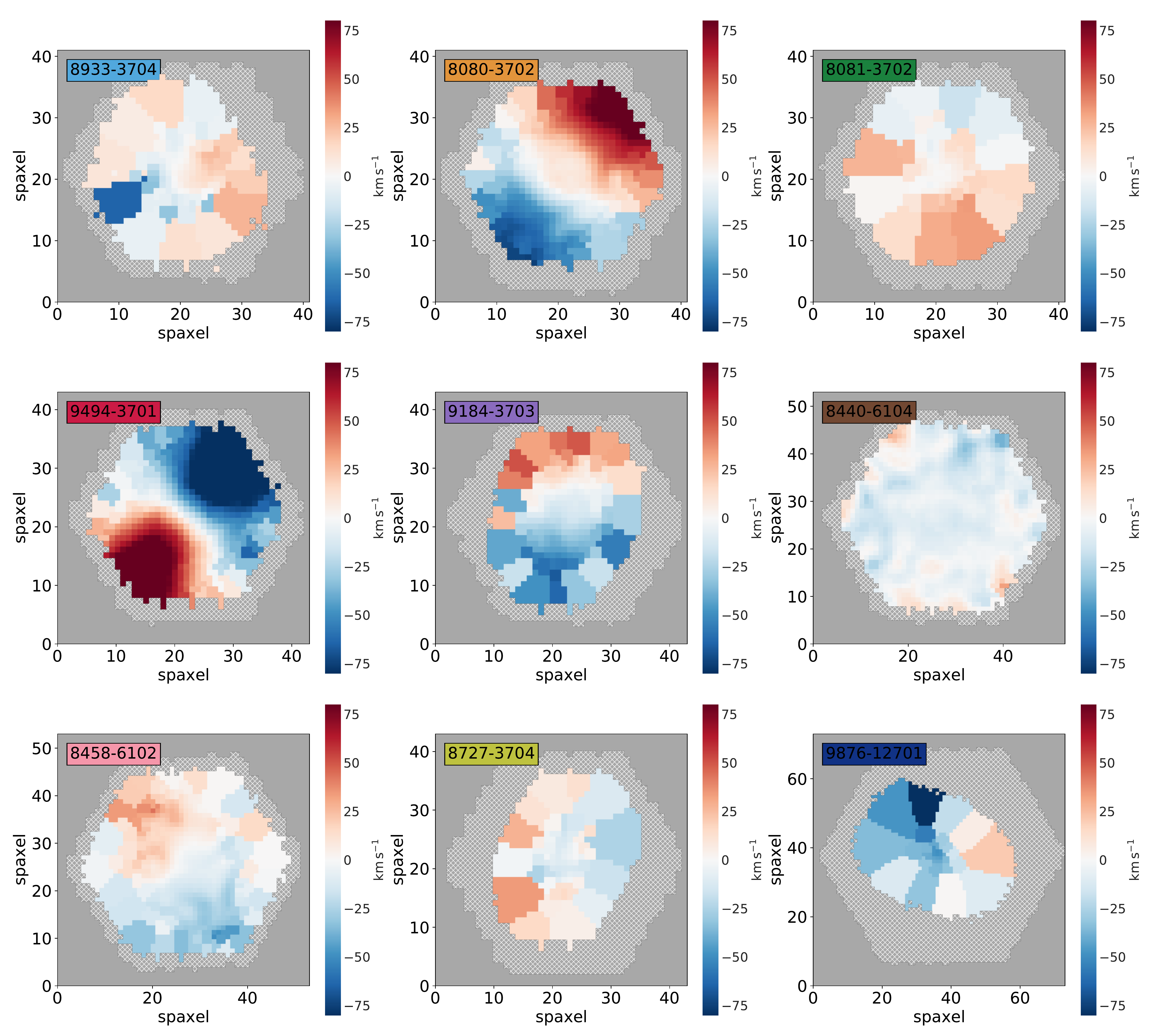}
	\caption{\edit1{The stellar velocity maps of post-starburst galaxies. The stellar velocities are taken from the MaNGA DAP with Voronoi bins. Visual inspection suggests that about half of the post-starburst galaxies have stellar velocity gradients along the major axes, suggesting rotational-supported kinematics.} \label{fig:stellar_vel}}
\end{figure*}

\begin{figure*}
	\centering
	\includegraphics[width=0.95\textwidth]{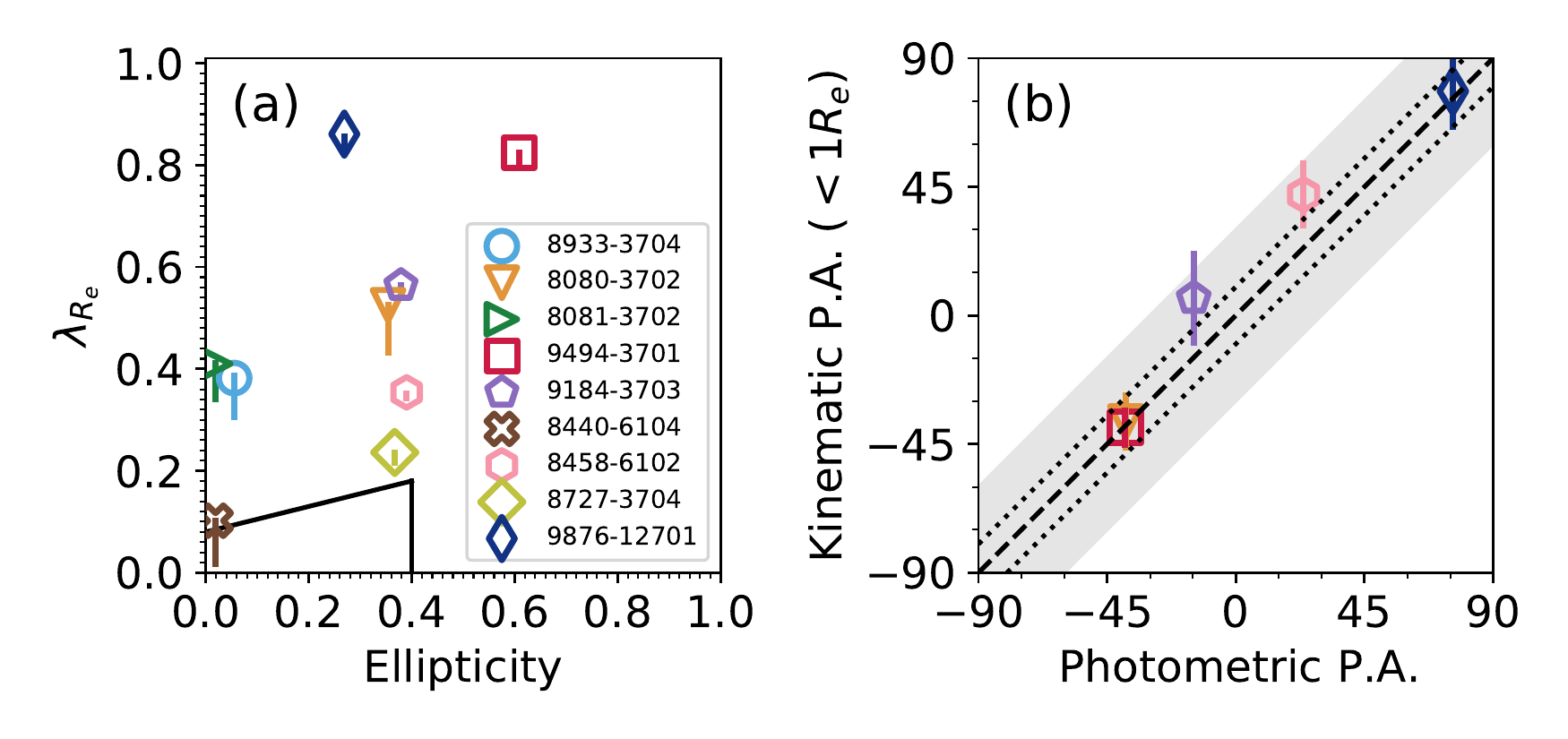}
	\caption{(a) The $\lambda_{R_e}$ and ellipticity ($1-b/a$) of post-starburst galaxies. The $\lambda_{R_e}$ is corrected for the FWHM of the data using the empirical prescription in \citet{gra18}. The wedge at the lower-left corner is the demacartion between fast and slow rotators. All post-starburst galaxies would be classified as fast rotators (b) The position angles of the major axis and the kinematic position angle, the direction of the maximum velocity measured within 1 $R_e$. The kinematic position angles and the 1-$\sigma$ uncertainties are measured using the python routin \texttt{fit\_kinematic\_pa} (see text). Galaxies with kinematic P.A. uncertainties larger than 30$\degree$ are not shown. The dotted lines and shaded area label $\pm10\degree$ and $\pm30\degree$ from the one-to-one relation, respectively. The photometric and kinematic axis of most (all) post-starburst galaxies are consistent within 10\degree (30\degree). \label{fig:pa}}
\end{figure*}

\edit1{Fig.~\ref{fig:stellar_vel} shows the stellar velocity maps of post-starburst galaxies. Quite a few show clear velocity gradients along the major axes from visual inspection, suggesting rotational-supported kinematics. The presence of velocity gradients is in line with the structural parameters: many of them are flat galaxies (e.g., 8080-3702, 9494-3701, 9184-3703, and 8458-6102, see Fig.~\ref{fig:psb_class} and Table~\ref{tab:sample}).}

I measure the luminosity-weighted stellar angular momentum \citep{ems07}:
\begin{equation}
\lambda_R \equiv \frac{\langle R|V| \rangle}{\langle R \sqrt{V^2+\sigma^2} \rangle} = \frac{\Sigma F_n R_n |V_n| }{ \Sigma F_n R_n \sqrt{V^2_n + \sigma^2_n} },
\end{equation}
where the summation is calculated over pixels within a radius $R$. 

Figure~\ref{fig:pa}a shows the $\lambda_{R_e}$ and the ellipticities ($1-b/a$) of post-starburst galaxies. The $\lambda_{R_e}$ is measured within $R_e$ and applied an emipircal correction for beam smearing\footnote{\url{https://github.com/marktgraham/ lambdaR_e_calc}} \citep{gra18}. Post-starburst are in general fast rotators based on the classification of \citet{ems07}. 

I then measure the kinematic position angles, which are the direction of the maximum velocities, using the the python routin \texttt{fit\_kinematic\_pa} \footnote{\url{http://www-astro.physics.ox.ac.uk/~mxc/software/\#pafit}} based on the method decribed in \citet{kra06}. Fig.~\ref{fig:pa}b shows the kinematic position angles and the position angles of the major axes. The kinematic position angles of 4 galaxies are unconstrained. Three galaxies (8933-3704, 8081-3702, and 8440-6104) are nearly face-on ($b/a > 0.9$). Visual inspection on the velocity map of the last galaxy (8727-3704) confirms the lack of a clear velocity gradient in the central $R_e$. For the 5 galaxies with reasonable constraints, the kinematic axis and the photometric axis align reasonably well, mostly within in 10\degree and all within 30\degree. 

The $\lambda_{R_e}$, ellipticity, and the alignment between kinematic and photometric axis of post-starburst galaxies are similar to flat fast rotators \citep{gra18}. Extending the measurements from $<1R_e$ to $<2R_e$ (when possible) yields a qualitatively the same conclusions. 

%The routine implements the method decribed in the Appendix C of \citet{kra06}. For a given position angle, the routine generates a quadrant symmetrical velocity field by taking the average velocities of corresponding Voronoi bin in each quadrant and calculate the $\chi^2$ between the observed and symmetrized velocity fields. The rounine then searches for the position angle that minimizes the $\chi^2$ and determines the best-fit position angle and the corresponding 3$\sigma$ confidence level.

%Tabel~\ref{tab:pa} lists the global stellar kinematic position angles within 1.5$R_e$ and the postition angles of the major axes. The stellar kinematic position angles of 3 galaxies are unconstrained. Two galaxies (8081-3702 and 8440-6104) have $b/a = 0.98$. They are likely face-on thus the rotation cannot be measured. Visual inspection on the volecity map of the third galaxy (8727-3704) confirms the lack of a clear velocity gradient. 

%For other 6 galaxies with reasonable constraints, the kinematic axis and the photometric axis align reasonably well, mostly within 10$\deg$ (Fig.~\ref{fig:pa}), similar to flat fast rotators \citep{gra18}. I also estimate the kinematic position angle within a smaller aperture of 1$R_e$ and find consistent results. 

%The well alignment between the photometric and kinematic axis suggests that most of the post-starburst galaxies unlikely had major merger events in the recent past \citep{ben00}. I will discuss the formation mechanisms in Sec.~\ref{sec:??}.  

\section{Discussion}
\label{sec:dis}
The sample of post-starburst galaxies selected by integrated IFU spectra gives us the anatomical view of galaxies experiencing overall quenching. I start the discussion from summarizing what we learn from the IFU data and the implication to the quenching mechanisms, then proceed to the comparison to post-starburst galaxies in the distant Universe.

\subsection{An overview of the quenching process}

While the sample is selected based on the spectra integrated over the entire IFU, the spatially-resolved spectra show a high degree of similarity with small individual variations. 
Firstly, the strong H$\delta$ absorption measured from the integrated spectra is representative for the entire galaxies, not dominated by small localized regions. Even the `quiescent' regions at galaxy outskirts have relative high \ewhd\ close to the fiducial $EW(H\delta) = 4$\AA\ cut (Fig.~\ref{fig:ind_r}a) and can be produced only by rapidly decline of star-formation rate. 
Secondly, compact ($\lesssim$1~kpc) emission-line regions are common at the centers of post-starburst galaxies and most of them are star-forming regions. While the \ewha\ is high enough to put these central regions in the `star-forming' category, their strong H$\delta$ absorption indicates the star-formation rates should have also declined rapidly (Fig.~\ref{fig:hahd}). The elevated gas-phase metalicities suggests that these galaxies have \edit1{low star-formation rates \citep{man10}} and low gas fraction \citep{zah14} thus likely at the end of their star-formation processes. Therefore, the centers of post-starburst galaxies are also undergoing the quenching processes and catched right before the star-formation fully stops.

\citet{bel18} analyzed the \ewha\ profiles of main squence star-forming galaxies in the MaNGA survey and showed that at the stellar mass range of the post-starburst galaxies studied in the paper, the star-forming galaxies have on average flat \ewha\ gradients. If the progenitors of post-starburst galaxies are normal star-forming galaxies, the higher \ewha\ at the centers of post-starburst galaxies suggest that the centers have elevated star-formation activities before quenching. It is also possible that no elevated star-formation activities happen, the quenching simply starts from outskirts then proceeds to the centers. A careful job on constraining the star-formation histories from spectra will be needed to differentiate these two scenarios.

Although post-starburst galaxies are located generally in the green valley (Fig.~\ref{fig:select}b), they have distinct stellar age and star-formation gradients. Green valley galaxies have on average flat ($M_\ast<10^{10}M_\odot$) or centrally-suppressed \ewha\ profiles, which are similar to star-forming galaxies \citep{bel18}. The luminosity-weighted ages of green valley galaxies are lower at outskirts, also consistent with normal quiescent galaxies \citep{gon14}. These results suggest that the transition from star-forming to quiescent is a smooth process that the star formation starts to shut down form galaxy centers. On the contrary, the peculiar star formation and stellar age gradients of post-starburst galaxies suggest that they have experienced different quenching mechanisms from normal green valley galaxies; there should be multiple quenching mechanisms at work. I will discuss more on the driving mechanisms in the next section. 

From 8 post-starburst galaxies with measureable emission lines, only 2 galaxies (8080-3702 and 9494-3701) have their centers dominated by AGN. These 2 galaxies also have low \ewha\ at the centers: $\sim$3\AA\ compared to $\sim10$\AA\ of other star-forming centers (Fig.~\ref{fig:ind_r}c). The absence of strong AGN is imposed by the selection criteria ($EW(H\alpha)_{em} \geq -3$\AA) thus should not be interpreted as a counterevidence against the presence of AGN in the quenching process. To understand the role of AGN, the upper limit on emission line strengths should be released \citep[see][]{yes14}. However, the absence of strong AGN and the residual star formation suggest that if AGN had involved in the quenching process earlier, it should have faded before the star formation fully stopped. In other words, the AGN does not shut down star-formation activities completely. 

\subsection{Quenching mechanisms}
\label{sec:mec}

The positive age gradients and quenching of the entire galaxies favors scenarios that involve some sorts of violent events that trigger gravitational instabilities. Gas loses angular momentum and collapses to the centers rapidly thus the star formation is stronger or lasts longer at centers. The inflowing gas is consumed and can also trigger AGN that heats up and expels gas in a short time so that the star formation stops quickly \citep{spr05,hop06}. 

Galaxy mergers have been long hypothesized as a plausible mechanism, as the mergers can achieve all these phenomena as well as transforming the morphology \citep{bar91,mih94,hop06,sny11}. Merger simulations suggest that gas-rich equal-mass mergers are the easiest combination to produce post-starburst galaxies as they can invoke instense starburst event then efficiently quench galaxies \citep{bek05,wil09,sny11,zhe20}. However, most post-starburst galaxies have their kinematic and photometric axes align with each other (Sec.~\ref{sec:kin}). The stellar kinematics disfavors equal-mass mergers as the main progenitors, which tend to create slow rotators with misaligned kinematic and photometric axes \citep{jes09}. 

If post-starburst galaxies are merger remanents, they are thus more likely produced by unequal-mass mergers. The loci on the \ewha---\ewhd\ plane (Fig.~\ref{fig:hahd}) also suggest that only a few percents of new stars are enough to produce post-starburst features. Strong starburst events invoked by equal-mass gas-rich mergers \edit1{are, while still possible,} not required to explain the spectral lines of most post-starburst galaxies in this paper. 
\edit1{In principle, the quality of the MaNGA spectra is good enough to perform full-spectral-fitting to constrain the star-formation histories of galaxies \citep[e.g.][]{san16,god17}. Knowing when the burst happened and the strength of the burst will further constrain the properties of the progenitors and the quenching mechanisms. However, the peculiar star-formation histories and stellar contents of post-starburst galaxies make the modeling a challenging task and the results and be model-dependent \citep[e.g., see][]{fre18}.}

I note that this merger scenario is likely not a complete picture for the formation of post-starburst galaxies. In the sample, galaxy 9876-12701 appears to be a special case. It has the lowest stellar mass, largest effective radius, bluest color, highest \ewhd\, and is the only galaxy without detectable H$\alpha$ emission. The merger scenario is not clearly applicable to galaxy 9876-12701. 

Moreover, with only 9 galaxies, the sample does not probe the environmental effect well. Two galaxies, 8933-3704 and 9876-12701, are in the Coma cluster, while others are either isolated or associated with small groups. While the formation mechanism of 9876-12701 is yet clear, the galaxy 8933-3704 fits into the merger scenario: a star-forming center and a negative \ewhd\ gradient. More cluster post-starburst galaxies are needed to understand whether and how cluster-specific mechanisms create post-starburst galaxies. 

\subsection{Comparison to high-redshifts}

While post-starburst galaxies are much more prevalent as redshifts approach unity and beyond \citep{yan09,wil16,row18a}, spatially-resolved information are spase at such redshifts. Previous spatially-resolved spectroscopic studies of post-starburst galaxies are mostly long-slit or IFU follow-up of relatively nearby galaxies ($z\lesssim0.1$) identified from fiber spectra \citep{pra05,got08a,got08b,pra10,pra13}, where the post-starburst features are measured from the very centers of galaxies.  

In this paper, I identify post-starburst galaxies from the spectra integrated over the entire IFUs. The large spectroscopic apertures make these galaxies more comparable to post-starburst galaxies at higher redshifts selected based on either spectral fetaures or SEDs measure from the bulk of the stellar components. Although there is no guarantee that these local and high-z post-starburst galaxies should have gone through the same process or be triggered by the same physical mechanisms, the spatially-resolved spectra are still highly informative to understand how galaxies with such peculiar observed features are produced. 

Several studies reported relatively small half-light radii of high-z post-starburst galaxies and in general attributed to \edit1{compact starburst events triggered by} highly-dissipative processes \citep{whi12,bel15,wu18a}. \citet{wu20} specifically proposed that the smaller half-light radii is a result of a centrally-concentrated starburst event on top of an aged population. \edit1{The most compact post-starburst galaxies ($R_e \simeq 1$~kpc) are those doubled the stellar masses in the starburst events. On the other hand,} even a small fraction of total mass, the young stars can make the light distribution in the optical wavelengths centrally concentrated. \edit1{In these cases, the half-light radii of post-starburst galaxies can be much smaller than their progenitor star-forming galaxies but not necessary smaller than typical quiescent galaxies \citep{wu18a,wu20}. Also, the half-mass radii are not necessarily significantly small \citep{sue20}.} The hypothesis is later strengthened by the discovery of weak negative \ewhd\ gradients\edit1{, which likely indicate positive age gradients,} from seeing-limited, deep slit spectra of post-starburst galaxies at $z\simeq0.8$ \citep{deu20}. \edit1{On the contrary, the age graidents of high-z ($z\sim2$) young quiescent galaxies are constrained for a couple of cases; A flat or a negative age gradient is seen in these galaxies \citep{akh20,jaf20}.} 

\edit1{The local post-starburst galaxies in this paper have qualitatively the same observed age gradients but from well resolved spectra. While the half-light radii of most local post-starburst galaxies are not much smaller than the average red galaxies, their moderate sizes are in line with a weak, centrally-enhanced starburst that added extra stellar light to the centers, therefore, the half-light radii are small compare to hypothesized progenitor star-forming galaxies. The lacking of extremely compact local star-forming galaxies may be explained by the average weaker burst strength. At $z\sim1$, post-starburst galaxies likely have doubled the stellar masses in the burst events thus the half-light radii can change significantly \citep{wil20,wu20}. On the contrary, if local post-starburst galaxies are the aftermath of weaker bursts (Fig.~\ref{fig:hahd}), possibly due to the lower gas fraction in galaxies at $z\sim0$, the half-light radii would not evolve as much. }

\edit1{While the central starburst scenario explains the positive age gradients}, not all high-z post-starburst galaxies studies find consistent \edit1{results}. \citet{set20} reported seeing-limited IFU observations of 6 massive ($M_\ast > 10^{11} M_\odot$), very young ($EW(H\delta) > 7\AA$) post-starburst galaxies at $z\sim0.6$. These galaxies have in general flat \ewhd\ gradients and are inconsistent with central compact starbursts on top of extended old stellar populations \citep[also see][]{hun18}. A possible coherent explanation is that unequal-mass mergers trigger galaxy-wide starburst and quenching processes, with stronger star-formation activities in galaxy centers as suggested by merger simulations \citep{zhe20}. These very young post-starburst galaxies are at the early phase of the quenching processes, where the outskirts have not aged to show a clear \ewhd\ gradient. 

Moreover, not all post-starburst galaxies have structures consistent with recent centrally-concnetrated star-formation events. \citet{mal18} reported that low-mass ($M_\ast \lesssim 10^{10} M_\odot$) post-starburst galaxies at $0.5<z<1.0$ have lower Sersic indice than quiescent galaxies of similar masses at four wavelengths from V to H, suggesting that these post-starburst galaxies do not have excess young stellar populations at their centers. Interestingly, the post-starburst galaxy with lowest stellar mass in this paper, 9876-12701, also has no sign of central yonger stellar population and is diffuse. Identifying more local dwarf post-starburst galaxies will help us to understand the mass dependence. 

\citet{mat20} reported that the half-light radii of post-starburst in $z\sim1$ clusters are between the star-forming and quiescent galaxies. Their toy models prefer a outside-in disk fading scenario for cluster post-starburst galaxies. Post-starburst galaxies in this paper do not sample the cluster environment well (Sec.~\ref{sec:mec}). Searching and resolving more local cluster post-starburst galaxies is needed for a definitve conclusion. 

In summary, the local post-starburst galaxies serves as a valuable proxy to understand the formation of their high-redshift counterparts. But the post-starburst galaxy population may be instrisically heterogenous, produced by different mechanisms under different conditions. If we would be able to identify the proper `counterparts' of high-redshift objects in the local Universe, the large angular extends of low-redshift galaxies will faciliate many observations that are not possible at high-redshifts. 

\section{Summary and future work}
\label{sec:sum}
Using the data from the SDSS MaNGA survey, I identify 9 galaxies whose spectra exhibit strong H$\delta$ absorption and weak H$\alpha$ emission integrated over the entire IFU plates. These galaxies will be identified as post-starburst galaxies if artifically shifted to high-redshifts and observed with typical observing facilities used for high-redshifts spectroscopy. They are thus the local counterparts of high-redshift post-starburst galaxies. The IFU spectra show following features. 

\begin{itemize}
	\item Almost all (8/9) local post-starburst galaxies have emission-line regions at their centers. Many of them would be classifed as emission-line galaxies based on fiber spectroscopy.
	\item Post-starburst galaxies have negative \ewhd\ gradients and positive \dn\ gradients, indicating that the stellar ages are younger at centers. The star-formation rates drop rapidly across the entire galaxy. 
	\item Only a small fraction (2/8) of the central emission line regions are dominated by AGN. The majority (6/8) is powered by star-forming activities within $\lesssim1$~kpc from the galaxy centers. The gas phase metalities is higher than the average mass-metallicity relation. 
	\item \edit1{A small fraction of young stars is enough to reproduce the \ewha\ and \ewhd\ of post-starburst galaxies.} 
	\item When measureable, the velocity gradients of the stellar components align with the photometric major axis. Their $\lambda_{R_e}$ and ellipticity classify them as fast rotators.
\end{itemize}

These observational features suggest that post-starburst galaxies in this paper are likely the products of unequal-mass gas-rich mergers that involke centrally-concentrated starburst events and galaxy-wide quenching, as described in merger simulations. This scenario can also explain the reported small half-light radii and the negative H$\delta$ gradients of post-starburst galaxies at $z\simeq1$ and beyond. The peculiar \ewha\ and \ewhd\ gradients suggest that the star formation in post-starburst galaxies and other green valley galaxies are shut down by different mechanisms. 

The sample is selected to have weak emission lines. The lack of AGN does not imply that AGN have no contribution to quenching. However, the prevalent residual star formation suggest that AGN does not fully shut down the star formation in the quenching process. To probe the role that AGN plays in the quenching process, other sample selections should be adopted. Furthermore, one post-starburst galaxy in the sample does not fit to the merger scenario, indicating that the post-starburst population may have heterogenous origins. Also, the sample does not probe the cluster environment well. Cluster-specific mechanisms are not tested in this paper. 

Because of their high surface brightness and large angular extends, local post-starburst galaxies are valuable proxies to collect observational information that would be difficult or impossible to obtain directly for high-redshifts galaxies. The currently-available and future IFU surveys on local galaxies are the repository for searching for these rare objects. 

\software{Astropy \citep{astropy2013,astropy2018}, pPXF \citep{cap04,cap17}, Marvin \citep{cher19}}

\acknowledgments
I thank the referee for constructive comments that makes this paper more complete. I thank the team that build the MaNGA interface Marvin, which makes the production of this paper much easier. I acknowledges the support of the fellowship from the East Asian Core Observatories Association. 

Funding for the Sloan Digital Sky Survey IV has been provided by the 
Alfred P. Sloan Foundation, the U.S. 
Department of Energy Office of 
Science, and the Participating 
Institutions. 

SDSS-IV acknowledges support and 
resources from the Center for High 
Performance Computing  at the 
University of Utah. The SDSS 
website is www.sdss.org.

SDSS-IV is managed by the 
Astrophysical Research Consortium 
for the Participating Institutions 
of the SDSS Collaboration including 
the Brazilian Participation Group, 
the Carnegie Institution for Science, 
Carnegie Mellon University, Center for 
Astrophysics | Harvard \& 
Smithsonian, the Chilean Participation 
Group, the French Participation Group, 
Instituto de Astrof\'isica de 
Canarias, The Johns Hopkins 
University, Kavli Institute for the 
Physics and Mathematics of the 
Universe (IPMU) / University of 
Tokyo, the Korean Participation Group, 
Lawrence Berkeley National Laboratory, 
Leibniz Institut f\"ur Astrophysik 
Potsdam (AIP),  Max-Planck-Institut 
f\"ur Astronomie (MPIA Heidelberg), 
Max-Planck-Institut f\"ur 
Astrophysik (MPA Garching), 
Max-Planck-Institut f\"ur 
Extraterrestrische Physik (MPE), 
National Astronomical Observatories of 
China, New Mexico State University, 
New York University, University of 
Notre Dame, Observat\'ario 
Nacional / MCTI, The Ohio State 
University, Pennsylvania State 
University, Shanghai 
Astronomical Observatory, United 
Kingdom Participation Group, 
Universidad Nacional Aut\'onoma 
de M\'exico, University of Arizona, 
University of Colorado Boulder, 
University of Oxford, University of 
Portsmouth, University of Utah, 
University of Virginia, University 
of Washington, University of 
Wisconsin, Vanderbilt University, 
and Yale University.

\bibliography{MaNGA_int_PSB_v3.2.bbl}
\end{document}